\documentclass[aps,prb,reprint,groupedaddress]{revtex4-2}
\usepackage{amsmath,amssymb,mathtools,bm}
\usepackage{graphicx}
\usepackage{booktabs,array,tabularx}
\usepackage{textcomp}
\usepackage{hyperref}
\hypersetup{unicode=true,colorlinks=true,linkcolor=blue,citecolor=blue,urlcolor=blue}
\graphicspath{{./Figures/}{./}}
\newcolumntype{C}{>{\centering\arraybackslash}X}
\allowdisplaybreaks
\newcommand{\dd}{\mathrm{d}}
\newcommand{\ee}{\mathrm{e}}
\newcommand{\ii}{\mathrm{i}}

\newcommand{\nth}{\bar n_{\mathrm{th}}}
\newcommand{\zpf}{x_{\mathrm{zpf}}}
\begin{document}
\title{Scanning Probe Control, Quantum Dynamics, and Dispersive Readout of Suspended Carbon Nanotube Resonators}
\author{Cristian Staii}
\email{cstaii01@tufts.edu}
\affiliation{Department of Physics and Astronomy, Tufts University, Medford, MA 02155, USA}
\date{\today}
\begin{abstract}

Suspended-carbon nanotubes combine small motional mass, high mechanical frequency, strong electrostatic tunability, and appreciable mechanical nonlinearity. These properties make them attractive for quantum nanomechanics, but they also place unusually stringent requirements on actuation and readout because a nearby control element can perturb the mechanical dynamics that it is intended to control or probe. We develop a unified analysis of a suspended-carbon nanotube (CNT) driven by a voltage-biased atomic force microscope (AFM) tip and coupled capacitively to a Cooper-pair box (CPB). First, we derive the driven linear and Duffing response of the CNT and identify how the static AFM tip force, force gradient, tip-position noise, and voltage noise enter the mechanical frequency, dephasing, and heating rates. This establishes explicit inequalities that quantify the trade-off between control strength and actuation-induced backaction during pulsed operation. Second, we quantize the Duffing mode, derive the Kerr coefficient using a fixed convention for the quartic potential, and connect spectral selectivity to drive bandwidth and leakage. Thermal occupations and active cooling estimates are given for representative nanotube modes from 54 MHz to 5.37 GHz. In the resolved-sideband limit, the required cooling cooperativity is modest for a 221 MHz mode but substantially larger for a 54 MHz mode at 10 mK. Third, we derive the CNT-CPB coupling from displacement-dependent gate capacitance, transform the charge-basis interaction to the qubit eigenbasis, and obtain the Jaynes--Cummings and dispersive Hamiltonians by controlled approximations. We then combine Ramsey phase accumulation that depends on the phonon number with quantum-capacitance readout of the CPB. The resulting framework identifies three experimentally distinct requirements: sufficiently weak AFM backaction during control, sufficiently strong mechanical anharmonicity for spectral selectivity, and sufficiently rapid dispersive mapping and superconducting readout for state-resolved detection. These results provide a quantitative operating framework for integrating local scanning-probe actuation with superconducting readout in CNT quantum electromechanics.
\end{abstract}
\maketitle

\section{Introduction}
\label{sec:introduction}

Nanoelectromechanical resonators provide a direct route to studying mechanical motion in regimes where zero-point fluctuations, thermal noise, nonlinear dynamics, and quantum measurement can all become experimentally relevant \cite{SchwabRoukes2005,LaHaye2004,OConnell2010,MacCabe2020, CirioPhysRevLett.109.147206}. Suspended-carbon nanotubes (CNTs) are especially favorable in this setting because their extremely small mass produces relatively large zero-point displacement, while electrostatic gates provide strong control over tension and resonance frequency \cite{Sazonova2004,Huttel2009,Steele2009,Schneider2014, Staii2005SGM, Staii2026Entropy}. Experiments have demonstrated high quality factors, strong coupling between charge motion and nanotube vibrations, nonlinear damping, electrostatic softening, cooling, and strongly nonlinear behavior approaching the quantum regime \cite{Huttel2009,milkie_controlled_switching_2005, Staii2005DNA,SilioStaii2026, Eichler2011,Urgell2020,Samanta2023}. These results place suspended CNTs among the most tunable solid-state mechanical systems available for mesoscopic quantum experiments. Although several elements of the analysis apply more generally to doubly clamped nanomechanical beams, CNTs provide a particularly favorable combination of properties for the present architecture. Their exceptionally small effective mass produces relatively large zero-point motion and correspondingly reduces the force required for coherent mechanical control, while strong electrostatic tunability and pronounced mechanical nonlinearity allow the resonance frequency and effective anharmonicity to be tuned over a broad range. In addition, their conductive nanoscale geometry supports displacement-dependent capacitive coupling both to the local AFM tip and to the CPB gate \cite{Sazonova2004,Huttel2009,Steele2009,Eichler2011,Samanta2023}. The framework is applicable more generally to other nanomechanical resonators, with the achievable operating regime determined by their mass, nonlinearity, coherence, tunability, and capacitive coupling.

Mechanical nonlinearity is central to this capability. For a doubly clamped resonator, transverse displacement stretches the beam and generates a geometric Duffing term even when the underlying elastic material is treated linearly \cite{LifshitzCross2009}. Additional nonlinearities arise from nearby electrostatic potentials and from amplitude-dependent damping \cite{LifshitzCross2009,Rhoads2010,Eichler2011,Kozinsky2006}. In the classical description, these effects produce amplitude-dependent resonance shifts, distorted response curves, and, at sufficiently strong drive, bistability \cite{BadzeyMohanty2005,Kozinsky2006}. In the quantum description, the same quartic potential produces a Kerr anharmonicity that makes adjacent vibrational transitions nondegenerate \cite{Semiao2009,RipsHartmann2013,Rips2014, Pistolesi2021}. A sufficiently large anharmonic separation allows the lowest two vibrational states to be addressed selectively and motivates proposals for nanomechanical qubits, phonon blockade, and nonclassical mechanical-state preparation \cite{lahaye_nanomechanical_2009, Semiao2009,RipsHartmann2013,Rips2014,Wang2016,Sarma2018}. The first requirement for using this anharmonicity for quantum state control is therefore that the separation between adjacent vibrational transitions be sufficiently large compared with the relevant spectral linewidths and drive bandwidth.

A second requirement for quantum control of the CNT mechanical mode is controlled actuation. In our previous work, we proposed using a nearby atomic force microscope (AFM) tip to apply calibrated local forces to a suspended CNT and thereby perform coherent control and quantum tomography of its fundamental flexural mode \cite{Chang2026}. The key experimental distinction is that the AFM cantilever need not oscillate at the CNT frequency. The cantilever and positioning stack establish a stable nanometer-scale working separation, while an electrical modulation of the tip bias, or an equivalent local gate voltage, produces a force component at the CNT resonance. This approach avoids optical heating and provides spatially localized actuation, but it raises a basic measurement question: how large can the useful resonant force be before the same tip produces unacceptable static softening, frequency noise, heating, or dephasing?

A quantitative treatment therefore requires viewing the AFM not simply as an ideal force source but as part of the electromechanical environment. A static AFM tip--CNT force shifts the equilibrium position, and its spatial derivative changes the effective spring constant \cite{GarciaSanchez2007}. Fluctuations of the tip--CNT gap convert into fluctuations of the resonance frequency, while voltage fluctuations convert into both low-frequency spring constant noise and force noise near the mechanical resonance. These effects are closely related to the general backaction constraints of displacement detection in quantum electromechanical systems \cite{Blencowe2004,Giessibl2003}. In a nonlinear CNT, these backaction and noise mechanisms are particularly important because frequency fluctuations broaden the mechanical transition through dephasing \cite{Schneider2014}, while resolving the Kerr anharmonicity requires the separation between adjacent vibrational transitions to exceed their spectral linewidths \cite{Rips2014,Pistolesi2021}.

\begin{figure}[t]

\centering
\includegraphics[width=0.95\columnwidth]{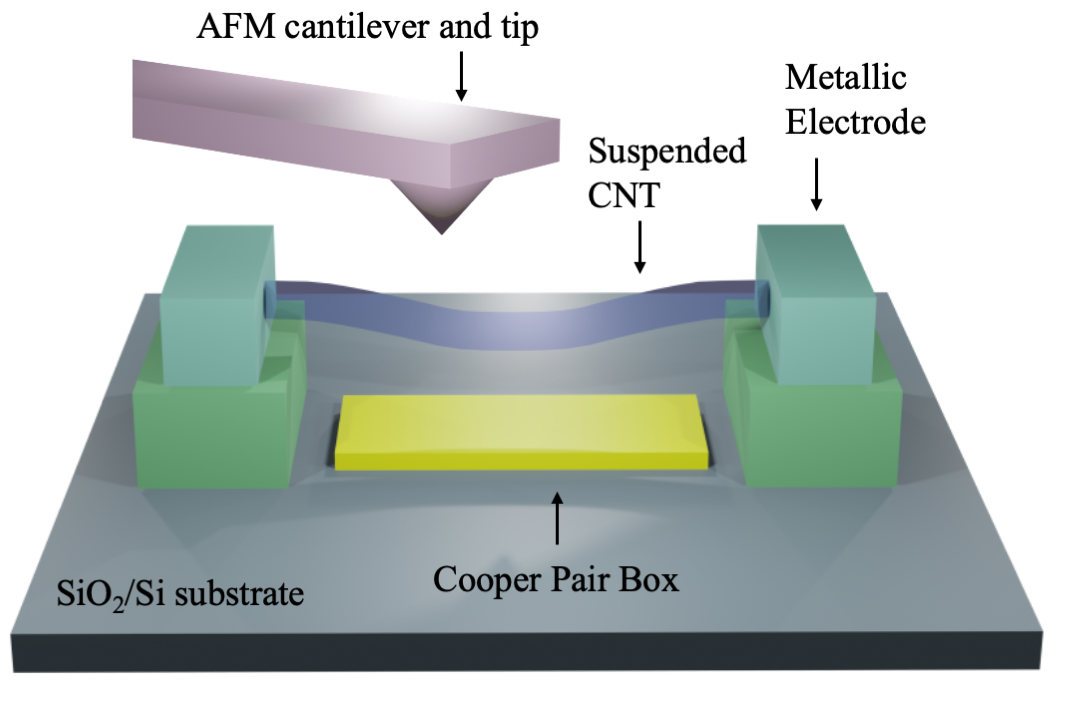}
\caption{Schematic of the proposed AFM-CNT-CPB quantum electromechanical platform. A doubly clamped single-walled carbon nanotube (CNT) of length $L=100$--$1000$ nm is suspended across a 200 nm trench and locally driven by an atomic force microscope (AFM) tip positioned near the nanotube. Operation at dilution refrigerator temperatures, $T\sim10$ mK, enables coherent control of the CNT flexural mode and Ramsey-based protocols for mapping mechanical-state information onto a superconducting two-level system. A Cooper-pair box (CPB) fabricated on the SiO$_2$/Si substrate is shown as a representative superconducting element for coupling to the CNT motion and for subsequent state-sensitive readout.}
\label{fig:setup}
\end{figure}

A third requirement is state-resolved readout. Superconducting circuits provide a natural platform because mechanical displacement can modulate the gate charge of a Cooper-pair box (CPB), producing an electromechanical interaction that is longitudinal in the charge basis and becomes predominantly transverse near the charge-degeneracy point after transformation to the qubit eigenbasis \cite{Armour2002,Blencowe2004,Pirkkalainen2013,pirkkalainen_cavity_2015,Lee2023}. In the near-resonant regime, this interaction leads to a Jaynes--Cummings Hamiltonian. In the large-detuning regime, it produces a dispersive interaction in which the CPB transition frequency acquires a shift that depends on the mechanical phonon number \cite{Blais2004,Zueco2009,Blais2021}. A Ramsey sequence can then convert this phonon number-dependent phase accumulation into a measurable CPB population. In our previous work, we showed that, for parity-sensitive tomography, choosing the interaction time so that the Ramsey phase changes by $\pi$ for each additional phonon maps the mechanical phonon parity onto the superconducting two-level system \cite{Chang2026}. The proposed experimental architecture is illustrated in Figure \ref{fig:setup}, which combines local electrical actuation of a suspended CNT by a voltage-biased AFM tip with capacitive coupling to a CPB for dispersive mechanical-state readout.

Once the mechanical-state information has been mapped onto the CPB, the remaining task is to measure the CPB state rapidly and with high fidelity. Several readout strategies are available, including cavity-based electromechanical detection and gate-based superconducting readout \cite{Regal2008,Teufel2011Strong}. Of particular relevance here is quantum-capacitance readout. The curvature of the CPB energy bands with respect to gate charge gives a state-dependent effective capacitance. Embedding the CPB in a radio-frequency resonator converts that capacitance into a resonance frequency and reflected-phase shift. Quantum-capacitance readout of superconducting charge devices has been demonstrated experimentally in several implementations \cite{Duty2005,Sillanpaa2005,Persson2010}. In particular, Persson and collaborators reported state discrimination on a time scale of approximately $50$--$100$ ns using a resonator with a quality factor of order $10^2$ \cite{Persson2010}. That experiment measured the parity state of the superconducting island associated with quasiparticle tunneling. In the CNT protocol considered here, the role is different: the CNT phonon parity is first mapped coherently onto the CPB state, after which quantum capacitance provides a fast electrical measurement of that CPB state. Keeping these two meanings of parity distinct is essential.

The central contribution of this work is a unified quantitative framework for determining whether local AFM actuation, nonlinear CNT dynamics, active cooling, and superconducting dispersive readout can operate compatibly within a single quantum electromechanical device. Rather than treating control, backaction, anharmonicity, cooling, and readout as independent elements, we connect them through experimentally accessible parameters and identify the quantitative constraints, rates, and time scales that define a viable operating regime. The resulting analysis establishes when an AFM tip can provide sufficiently strong local control without introducing excessive frequency shifts, noise, heating, or dephasing, when the intrinsic CNT nonlinearity is large enough to resolve individual vibrational transitions, and when CNT-to-CPB coupling can support phonon-sensitive dispersive mapping followed by rapid electrical readout.

This unified approach is relevant to several areas of condensed matter and quantum science. Suspended CNTs provide a particularly well-controlled setting in which mechanical nonlinearity, mesoscopic electromechanical coupling, environmental noise, and superconducting quantum circuitry can be studied within the same solid-state platform. The analysis therefore connects nanoscale condensed matter dynamics with questions of decoherence and open quantum system behavior \cite{schlosshauer_decoherence_2007, breuer_theory_2009}, including how local control fields and fluctuating electromagnetic environments modify mechanical coherence and quantum state resolution. At the same time, the combination of anharmonic mechanical states, coherent electromechanical coupling, and dispersive superconducting readout provides a route toward phonon-resolved measurements and mechanically encoded quantum information. These connections make the proposed architecture relevant not only to quantum nanomechanics, but also to hybrid quantum systems, quantum sensing, and the development of mechanically based elements for quantum information processing.

The paper is organized as follows. Section \ref{sec:driven} derives the AFM-driven linear and Duffing response of the CNT, including the relation between the applied tip voltage and the resonant force. Section \ref{sec:backaction} identifies the principal AFM-induced backaction channels and formulates quantitative constraints on stable and coherent operation. Section \ref{sec:kerr} connects the classical Duffing nonlinearity to the quantum Kerr Hamiltonian and develops active cooling estimates for representative CNT modes. Section \ref{sec:cpb} derives the CNT--CPB interaction from circuit electrostatics, obtains the near-resonant and dispersive limits, and analyzes Ramsey-based phonon number mapping followed by quantum-capacitance readout. Section \ref{sec:window} combines these results into an experimentally useful hierarchy of rates, energy scales, and time scales and outlines a practical calibration sequence. Section \ref{sec:discussion} discusses the resulting operating constraints, experimental limitations, and the broader relevance of the proposed architecture to mesoscopic physics, decoherence studies, and hybrid quantum systems. Finally, Section \ref{sec:conclusion} summarizes the principal results and their implications. The overall objective is to establish the physical conditions under which localized AFM control, nonlinear CNT quantum dynamics, and superconducting state-resolved detection can be implemented simultaneously in the same device.

\section{AFM-Driven CNT: Linear and Nonlinear Response}
\label{sec:driven}

\subsection{Single-Mode Equation of Motion}

Near its fundamental flexural resonance, a doubly clamped nanomechanical resonator can be reduced to an effective single-mode Duffing oscillator \cite{LifshitzCross2009}. For carbon nanotube (CNT) resonators, both elastic nonlinearity and amplitude-dependent dissipation can become experimentally important \cite{LifshitzCross2009,Rhoads2010,Eichler2011}. We describe the fundamental flexural mode by a generalized coordinate $q(t)$, measured relative to the static equilibrium configuration corresponding to the chosen AFM tip position and applied dc bias. In a single-mode approximation, the transverse displacement of the CNT can be written as $u(x,t)=\phi(x)q(t)$, where $u(x,t)$ is the transverse displacement at position $x$ along the nanotube and $\phi(x)$ is the spatial profile of the fundamental flexural mode \cite{LifshitzCross2009,Rhoads2010}. We choose the normalization $\phi(x_{\mathrm{tip}})=1$, where $x_{\mathrm{tip}}$ is the position at which the AFM force is applied. With this convention, $q(t)$ is equal to the physical displacement of the CNT at the position $x_{\mathrm{tip}}$ where the AFM force is applied. For a different mode normalization, the externally applied force must be projected onto the mechanical mode. A localized force $F(t)$ acting at $x_{\mathrm{tip}}$ then enters the single-mode equation through the generalized force $F_{\mathrm{mode}}(t)=\phi(x_{\mathrm{tip}})F(t)$ \cite{LifshitzCross2009,Rhoads2010,Eichler2011}. Let $m_{\mathrm{eff}}$ denote the effective mass of the mode, $\omega_m$ its small-amplitude angular resonance frequency at the chosen static working point, and $\gamma_m$ its linear energy damping rate. We use the convention $Q_m=\omega_m/\gamma_m$, where $Q_m$ is the small-amplitude mechanical quality factor. We denote the cubic coefficient of the restoring force by $\alpha$ and the leading nonlinear damping coefficient by $\eta$. The resonant drive has generalized force amplitude $F_d$, angular frequency $\omega_d$, phase $\varphi_d$, and dimensionless envelope $s(t)$. Here $s(t)$ represents the externally prescribed temporal envelope of the resonant electrical drive. It controls how the drive is turned on, turned off, or shaped during a finite control pulse, with $s(t)=1$ corresponding to continuous-wave excitation. The envelope is assumed to vary slowly compared with the carrier oscillation at $\omega_d$, so that $s(t)\cos(\omega_d t+\varphi_d)$ describes a gated or shaped near-resonant force. It is not associated with mechanical motion of the AFM cantilever or with stochastic voltage fluctuations. These effects are treated separately through the static tip position and the noise term $\delta F(t)$. In Section \ref{subsec:electrostatic_drive}, the same envelope is used in the applied tip voltage, and therefore directly determines the temporal profile of the resonant AFM force. Finally, let $\delta F(t)$ denote the zero-mean stochastic generalized force associated with the AFM and other environmental degrees of freedom. With these definitions, the effective equation of motion for the AFM-driven CNT is \cite{LifshitzCross2009,Rhoads2010,Eichler2011}
\begin{equation}
\begin{aligned}
&m_{\mathrm{eff}}\ddot q+m_{\mathrm{eff}}\gamma_m\dot q
+m_{\mathrm{eff}}\omega_m^2q\\
&\quad+\alpha q^3+\eta q^2\dot q\\
&=F_d s(t)\cos(\omega_d t+\varphi_d)+\delta F(t).
\end{aligned}
\label{eq:duffing_eom}
\end{equation}

The terms proportional to $\alpha$
 and $\eta$ describe, respectively, the leading Duffing restoring nonlinearity and amplitude-dependent damping. The latter has been observed to be particularly important in CNT and graphene resonators \cite{Eichler2011}. Neglecting damping, driving, and fluctuations, the conservative part of Equation \eqref{eq:duffing_eom} follows from the effective potential \cite{LifshitzCross2009,Rhoads2010}
\begin{equation}
 U(q)=
 \frac{1}{2}m_{\mathrm{eff}}\omega_m^2q^2
 +\frac{\alpha}{4}q^4,
\label{eq:duffing_potential}
\end{equation}
where $U(q)$ is the mechanical potential energy. A positive Duffing coefficient, $\alpha>0$, produces a hardening response, whereas $\alpha<0$ produces a softening response. For $\alpha<0$, Equation \eqref{eq:duffing_potential} should be interpreted as a local low-amplitude expansion because higher-order restoring terms must ultimately stabilize the physical system.

Equation \eqref{eq:duffing_eom} retains the dominant symmetric Duffing nonlinearity. In general, expanding the electrostatic interaction about a biased equilibrium can also generate quadratic and higher-order terms in the restoring force \cite{LifshitzCross2009,Rhoads2010}. For the operating regime considered here, however, the CNT motion remains at sufficiently small amplitude that these additional contributions are expected to be weak compared with the effective cubic nonlinearity. Neglecting them therefore provides a good approximation for the present system while retaining the leading nonlinear behavior relevant to the driven response and Kerr anharmonicity. The parameters $\omega_m$ and $\alpha$ are accordingly understood as effective quantities evaluated at the specified AFM tip position and dc bias.

\subsection{Electrostatic Actuation by the AFM Tip}
\label{subsec:electrostatic_drive}

Electrostatic actuation of CNT and other nanomechanical resonators can be described in terms of the displacement dependence of the AFM tip--resonator capacitance \cite{Sazonova2004,GarciaSanchez2007,Giessibl2003}. Let $C_t(z)$ be the effective capacitance between the voltage-biased AFM tip and the CNT, where $z$ is the local tip--CNT separation. We choose positive $q$ toward the tip, so that locally $z=z_0-q$, with $z_0$ the static working separation. At fixed applied voltage $V(t)$, the electrostatic generalized force is obtained from the capacitance gradient \cite{Giessibl2003,GarciaSanchez2007},
\begin{equation}
 F_{\mathrm{el}}(q,t)
 =
 \frac{1}{2}\frac{\partial C_t}{\partial q}V(t)^2
 =
 -\frac{1}{2}\frac{\partial C_t}{\partial z}V(t)^2.
\label{eq:electrostatic_force}
\end{equation}

Let
 $V_{\mathrm{dc}}$ denote the effective static voltage difference between the AFM tip and the CNT at the working point, including any static contact-potential offset, and let $V_{\mathrm{ac}}$ denote the amplitude of the radio-frequency modulation of this voltage difference. In the configuration considered here, the modulation is applied to the conductive AFM tip while the CNT is held at a fixed electrical potential. The AFM cantilever itself remains mechanically quasi-static. Using the same drive frequency $\omega_d$, phase $\varphi_d$, and envelope $s(t)$ introduced above, we write
\begin{equation}
 V(t)=
 V_{\mathrm{dc}}
 +V_{\mathrm{ac}}s(t)\cos(\omega_d t+\varphi_d).
\label{eq:tip_voltage}
\end{equation}

Squaring Equation \eqref{eq:tip_voltage} produces a static component, a component at $\omega_d$, and a component at $2\omega_d$. For $V_{\mathrm{ac}}\ll V_{\mathrm{dc}}$, the component at $\omega_d$ is dominated by the cross term proportional to $V_{\mathrm{dc}}V_{\mathrm{ac}}$. The corresponding resonant-force amplitude is therefore \cite{Sazonova2004,GarciaSanchez2007,GarciaSanchez2008}
\begin{equation}
 F_d\simeq
 \left|
 \frac{\partial C_t}{\partial z}
 \right|
 V_{\mathrm{dc}}V_{\mathrm{ac}}.
\label{eq:drive_force}
\end{equation}
If the mode coordinate is not normalized to the displacement at the tip position, the right-hand side of Equation \eqref{eq:drive_force} must also be multiplied by the corresponding mode-shape factor $\phi(x_{\mathrm{tip}})$.

An important consequence of Equation \eqref{eq:drive_force} is that the AFM cantilever itself does not have to oscillate at the CNT resonance. The cantilever and positioning system can establish the quasi-static AFM tip--CNT separation, while the electrical modulation generates the high-frequency force. Consequently, the mechanical bandwidth of the AFM cantilever does not determine the carrier frequency of the CNT drive. The required bandwidth must instead be supported by the electrical wiring and the tip--CNT circuit. Related electrical and scanning probe approaches have been used to actuate and detect nanoscale mechanical motion \cite{GarciaSanchez2007,GarciaSanchez2008,SanPaulo2007,GarciaHerruzo2012}.

For an order-of-magnitude estimate, we approximate the local tip--CNT capacitance by an effective area $A_{\mathrm{eff}}$ separated from the CNT by $z_0$. Let $\epsilon_0$ denote the vacuum permittivity. Taking $A_{\mathrm{eff}}\sim(50\,\mathrm{nm})^2$ and $z_0\sim50\,\mathrm{nm}$ gives the approximate capacitance gradient used in Ref. \cite{Chang2026},
\begin{equation}
 \left|
 \frac{\partial C_t}{\partial z}
 \right|
 \sim
 \frac{\epsilon_0 A_{\mathrm{eff}}}{z_0^2}
 \sim 10^{-11}\ \mathrm{F/m}.
\label{eq:cap_gradient_estimate}
\end{equation}
This estimate is intended only to set the force scale because the actual capacitance depends on the tip geometry and the surrounding electrostatic environment. For example, $V_{\mathrm{dc}}\sim0.1$ V and $V_{\mathrm{ac}}\sim1$ mV give $F_d\sim1$ fN, while smaller rf amplitudes give sub-fN forces. These values are consistent with the representative force scales estimated for coherent CNT control in Ref. \cite{Chang2026}. The capacitance-gradient formulation does not require a specific AFM tip shape. The apex radius, cone angle, and tip--CNT separation enter through $C_t(z)$ and its spatial derivatives and therefore determine both the resonant driving force and the static force-gradient backaction. For a sharp tip at nanometer-scale separation, the closest apex region generally provides the most localized and strongly distance-dependent contribution, while the conical body and cantilever can contribute a smoother background capacitance. For a particular device, quantitative predictions can be based on experimental calibration of $C_t(z)$ as a function of tip--CNT separation, from which the relevant capacitance derivatives can be determined \cite{Giessibl2003,GarciaSanchez2007,GarciaSanchez2008, Yurchenko2020}.

\subsection{Harmonic Response and Pulse Ring-Up}

The linear driven response provides the connection between the applied AFM force and the measured CNT displacement. The standard expressions for a damped driven resonator follow from Equation \eqref{eq:duffing_eom} by setting $\alpha=\eta=0$, neglecting $\delta F(t)$, and taking $s(t)=1$ for an indefinite continuous-wave (CW) drive \cite{LifshitzCross2009,Rhoads2010,Chang2026}. Let $A_{\mathrm{lin}}(\omega_d)$ be the steady-state displacement amplitude and $\phi_{\mathrm{lin}}$ the phase lag of the CNT response relative to the applied force. These quantities are given by the well-known expressions:
\begin{equation}
 A_{\mathrm{lin}}(\omega_d)=
 \frac{F_d/m_{\mathrm{eff}}}
 {\sqrt{
 (\omega_m^2-\omega_d^2)^2+
 \gamma_m^2\omega_d^2
 }},
\label{eq:linear_amplitude}
\end{equation}
and
\begin{equation}
 \tan\phi_{\mathrm{lin}}
 =
 \frac{\gamma_m\omega_d}
 {\omega_m^2-\omega_d^2},
\label{eq:linear_phase}
\end{equation}
where the appropriate quadrant of $\phi_{\mathrm{lin}}$ is chosen according to the sign of $\omega_m^2-\omega_d^2$. It is useful to introduce the dimensionless frequency ratio $r=\omega_d/\omega_m$ and damping ratio $\zeta=\gamma_m/(2\omega_m)=1/(2Q_m)$. Defining the dimensionless magnification factor $M(r,\zeta)$ as the driven displacement divided by the static compliance scale $F_d/(m_{\mathrm{eff}}\omega_m^2)$ gives \cite{LifshitzCross2009,Rhoads2010}
\begin{equation}
 M(r,\zeta)=
 \frac{1}
 {\sqrt{(1-r^2)^2+(2\zeta r)^2}}.
\label{eq:magnification}
\end{equation}
At $\omega_d=\omega_m$, Equation \eqref{eq:magnification} gives $M=Q_m$. For $Q_m\gg1$, this frequency is also very close to the frequency of maximum response. This resonant enhancement explains why femtonewton and sub-femtonewton forces can produce appreciable steady-state displacements in a high-$Q_m$ CNT.

The time-domain control protocols considered here do not generally operate in the infinite-time CW limit. For a finite drive pulse, the linear oscillator response can instead be expressed through its causal Green function \cite{Chang2026}. Define
\begin{equation}
 \omega_1=
 \sqrt{\omega_m^2-\frac{\gamma_m^2}{4}},
\end{equation}
as the damped angular frequency in the underdamped regime, and let $q_{\mathrm{hom}}(t)$ denote the homogeneous solution fixed by the initial displacement and velocity. The exact linear response to an arbitrary envelope $s(t)$ is then \cite{Arfken2012}
\begin{equation}
\begin{aligned}
q(t)&=q_{\mathrm{hom}}(t)
+\frac{F_d}{m_{\mathrm{eff}}\omega_1}\int_0^t\dd t'\,\\
&\quad\times \ee^{-\gamma_m(t-t')/2}
\sin\!\left[\omega_1(t-t')\right]\\
&\quad\times s(t')\cos(\omega_dt'+\varphi_d).
\end{aligned}
\label{eq:transient_green}
\end{equation}
With the damping convention used here, the displacement amplitude decays with time constant $2/\gamma_m$, whereas the mechanical energy decays with time constant $1/\gamma_m$. The latter is the mechanical energy-relaxation time,
\begin{equation}
 T_1=\frac{1}{\gamma_m}
 =\frac{Q_m}{\omega_m}.
\label{eq:T1_mechanical}
\end{equation}
A gated carrier can therefore contain many oscillation cycles while remaining well within the transient ring-up regime.

As representative examples, consider $Q_m=10^4$ and ordinary mechanical frequencies $f_m=\omega_m/(2\pi)$ of $5.37$ GHz, $221$ MHz, and $54$ MHz. These values give $T_1\simeq0.29$, $7.4$, and $29.6$ $\mu$s
, respectively \cite{Chang2026}. They are representative model parameters rather than measurements of a single CNT device. For drive pulses of duration $t_p\simeq0.1$ $\mu$s at $5.37$ GHz and $t_p\simeq1$ $\mu$s at $221$ MHz and $54$ MHz, the corresponding ratios are approximately $t_p/T_1=0.34$, $0.14$, and $0.034$, respectively. At the same time, these pulses contain approximately $537$, $221$, and $54$ carrier cycles. They can therefore be treated as near-resonant gated bursts rather than as impulsive mechanical kicks.

\subsection{Duffing Response Under Continuous Drive}

The nonlinear steady-state response of a directly driven Duffing resonator can be obtained by harmonic balance, retaining only the fundamental Fourier component of the response \cite{LifshitzCross2009,Rhoads2010,Eichler2011}. Let $A$ denote the positive oscillation amplitude and $\phi$ the phase lag relative to the drive. The leading single harmonic approximation is
\begin{equation}
 q(t)\simeq A\cos(\omega_dt-\phi).
\label{eq:duffing_ansatz}
\end{equation}
For this ansatz, the fundamental component of the cubic restoring term
is $q^3\simeq(3/4)A^3\cos(\omega_dt-\phi)$, whereas the fundamental component of the nonlinear damping term is $q^2\dot q\simeq-(1/4)A^3\omega_d \sin(\omega_dt-\phi)$. Substituting these components into Equation \eqref{eq:duffing_eom} and neglecting the stochastic force gives the harmonic-balance amplitude relation \cite{LifshitzCross2009,Rhoads2010,Eichler2011}
\begin{equation}
\begin{aligned}
\left(\frac{F_d}{m_{\mathrm{eff}}}\right)^2
&\simeq A^2\bigg[\bigg(\omega_m^2-\omega_d^2
+\frac{3\alpha A^2}{4m_{\mathrm{eff}}}\bigg)^2\\
&\qquad+\omega_d^2\bigg(\gamma_m
+\frac{\eta A^2}{4m_{\mathrm{eff}}}\bigg)^2\bigg].
\end{aligned}
\label{eq:duffing_response}
\end{equation}
The corresponding phase relation is
\begin{equation}
 \tan\phi
 \simeq
 \frac{
 \omega_d\left(
 \gamma_m+\eta A^2/(4m_{\mathrm{eff}})
 \right)
 }{
 \omega_m^2-\omega_d^2+
 3\alpha A^2/(4m_{\mathrm{eff}})
 }.
\label{eq:duffing_phase}
\end{equation}

\begin{figure*}[t]

\centering
\includegraphics[width=0.94\textwidth]{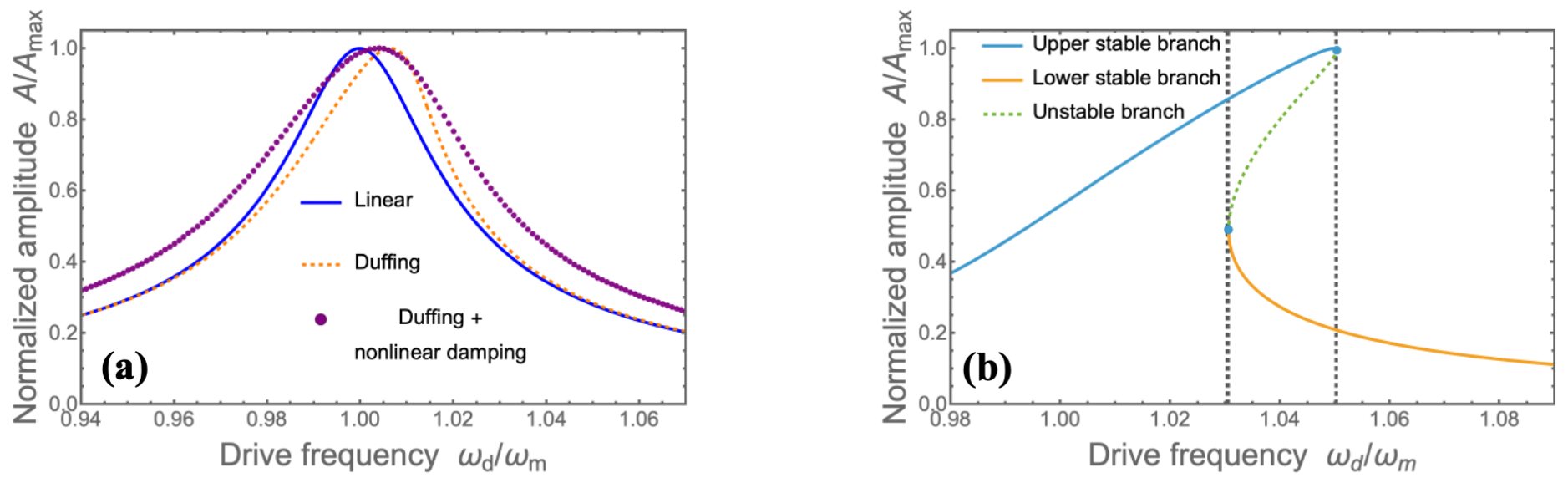}
\caption{Nonlinear response of the driven Duffing resonator. (\textbf{a}) Normalized steady-state amplitude $A/A_{\max}$ as a function of the normalized drive frequency $\omega_d/\omega_m$ for the linear response (continuous blue), the Duffing response (dashed orange), and the Duffing response including nonlinear damping (purple circles). Each curve is normalized to its own maximum amplitude.
(\textbf{b}) Normalized Duffing response at stronger drive, showing the upper (blue) and lower (orange) stable branches, the unstable intermediate branch (dashed green), and the fold points delimiting the bistable region, marked by blue circles. The vertical dashed lines indicate the corresponding fold frequencies and delimit the frequency interval over which the bistable, and therefore hysteretic, response occurs. In this panel, all amplitudes are normalized by the same value $A_{\max}$. Both panels use dimensionless parameters with $m_{\mathrm{eff}}=\omega_m=1$ and are intended to illustrate the generic nonlinear response.}
\label{fig:duffing_response}
\end{figure*}

Equations \eqref{eq:duffing_response} and \eqref{eq:duffing_phase} show explicitly that the Duffing nonlinearity produces an amplitude-dependent resonance shift, while nonlinear damping introduces an amplitude-dependent contribution to the effective linewidth. The consequences of these nonlinear terms are illustrated in Figure \ref{fig:duffing_response}. Figure \ref{fig:duffing_response}a compares the linear response with that obtained for a positive Duffing nonlinearity and for the combined effects of Duffing nonlinearity and nonlinear damping. The Duffing term shifts and distorts the resonance, whereas nonlinear damping modifies its linewidth and line shape. At stronger drive, the steady-state response can become multivalued, as shown in Figure \ref{fig:duffing_response}b. Over a finite frequency interval, two dynamically stable oscillation amplitudes coexist and are separated by an unstable solution (dotted green line). The boundaries of this bistable region correspond to the fold points of the implicit amplitude-frequency response and satisfy $d\omega_d/dA=0$ along the steady-state response curve. The parameters used in Figure \ref{fig:duffing_response} are chosen to display these characteristic features clearly and do not represent a quantitative fit to a specific CNT resonator. The bistable structure shown in Figure \ref{fig:duffing_response}b is characteristic of a weakly damped Duffing resonator \cite{LifshitzCross2009,Kozinsky2006,BadzeyMohanty2005}. During a slow frequency sweep, the oscillator follows one stable branch until a fold point is reached (blue circle in Figure \ref{fig:duffing_response}b), at which point it switches discontinuously to the other stable branch, producing hysteresis. The vertical dashed lines in Figure \ref{fig:duffing_response}b mark the two fold frequencies and define the hysteretic frequency interval.

Experimental measurements of the nonlinear amplitude and phase response can be used to extract the effective Duffing coefficient $\alpha$ and, when relevant, the nonlinear damping coefficient $\eta$, while also identifying the drive range over which a linear transfer function remains adequate \cite{LifshitzCross2009,Eichler2011}. For quantum control, however, operation deep in the classical bistable regime is not required. The relevant quantity is instead the anharmonic separation between adjacent quantum transitions, derived in Section \ref{sec:kerr}. Let $|0\rangle$, $|1\rangle$, and $|2\rangle$ denote the lowest vibrational number states of the CNT mode. Because of the Kerr anharmonicity, the transition frequency for $|1\rangle\rightarrow|2\rangle$ differs from that for $|0\rangle\rightarrow|1\rangle$ \cite{Chang2026}. We refer to this frequency difference as the detuning of the $|1\rangle\rightarrow|2\rangle$ transition relative to the fundamental $|0\rangle\rightarrow|1\rangle$ transition. Selective control of the lowest two states therefore requires the Rabi frequency and spectral bandwidth of the drive to remain smaller than this anharmonic detuning, while the corresponding classical displacement amplitude can remain well below the strongly nonlinear bistable regime.

\section{AFM--CNT Interaction and Backaction Constraints}
\label{sec:backaction}

\subsection{Static Force Gradient and Mechanical Stability}

The AFM tip modifies the CNT dynamics not only through the resonant force derived in Section \ref{sec:driven}, but also through the static spatial dependence of the tip--CNT interaction. The effect of the force gradient is well established in dynamic AFM and in electrostatically tuned nanomechanical systems \cite{Giessibl2003,RipsHartmann2013,Chang2026}. Let $F_{\mathrm{ts}}(q)$ denote the total static tip--sample force, including electrostatic and dispersion contributions. Expanding this force about the chosen static working point gives
\begin{equation}
\begin{aligned}
F_{\mathrm{ts}}(q)
&=F_{\mathrm{ts}}^{(0)}+G_F q
+\frac{1}{2}G_F^{(2)}q^2+\cdots,\\
G_F&=\left.\frac{\partial F_{\mathrm{ts}}}{\partial q}\right|_{q=0}.
\end{aligned}
\label{eq:force_gradient_expansion}
\end{equation}
where $F_{\mathrm{ts}}^{(0)}$ is the static tip force, $G_F$ is its linear force gradient, and $G_F^{(2)}=\left.\partial^2F_{\mathrm{ts}}/\partial q^2\right|_{q=0}$ is the next spatial derivative. Because $q=0$ was defined in Section \ref{sec:driven} as the static equilibrium position at the chosen tip position and dc bias, the constant term $F_{\mathrm{ts}}^{(0)}$ is already balanced by the static mechanical restoring force and any other static forces. The quadratic and higher terms are assumed to be small over the operating displacement range, consistent with the effective Duffing approximation introduced in the preceding section.

To distinguish the unperturbed mechanical stiffness from the stiffness in the presence of the local tip gradient, we define
\begin{equation}
 k_{m,0}=m_{\mathrm{eff}}\omega_{m,0}^2,
\end{equation}
where $\omega_{m,0}$ is the reference small-amplitude resonance frequency before the local AFM force gradient $G_F$ is included. The force gradient modifies the effective spring constant according to \cite{Giessibl2003,RipsHartmann2013}
\begin{equation}
\begin{aligned}
k_{\mathrm{eff}}
&=k_{m,0}-G_F\\
&=m_{\mathrm{eff}}\omega_{m,0}^2-G_F,\\
\omega_m^2
&=\omega_{m,0}^2-\frac{G_F}{m_{\mathrm{eff}}}.
\end{aligned}
\label{eq:gradient_frequency}
\end{equation}
The quantity $\omega_m$ in Equation \eqref{eq:gradient_frequency} is the same effective resonance frequency used throughout Section \ref{sec:driven}. For a weak force gradient, $|G_F|\ll k_{m,0}$, the corresponding shift $\delta\omega_m=\omega_m-\omega_{m,0}$ is
\begin{equation}
 \frac{\delta\omega_m}{\omega_{m,0}}
 \simeq
 -\frac{G_F}
 {2m_{\mathrm{eff}}\omega_{m,0}^2}
 =
 -\frac{G_F}{2k_{m,0}}.
\label{eq:frequency_shift_gradient}
\end{equation}
For the electrostatic interaction introduced in Section \ref{subsec:electrostatic_drive}, the static contribution to the force gradient scales approximately as $V_{\mathrm{dc}}^2$. In the mode normalization used here,
\begin{equation}
 G_F^{(\mathrm{el})}
 \simeq
 \frac{1}{2}V_{\mathrm{dc}}^2
 \left.
 \frac{\partial^2 C_t}{\partial q^2}
 \right|_{q=0},
\label{eq:electrostatic_gradient}
\end{equation}
where $C_t$ is the tip--CNT capacitance and $V_{\mathrm{dc}}$ includes the effective contact-potential offset introduced previously.

The exact static-stability requirement for the CNT mode is
\begin{equation}
 k_{\mathrm{eff}}>0.
\label{eq:exact_cnt_stability}
\end{equation}
Condition  \eqref{eq:exact_cnt_stability} ensures that the equilibrium remains a local minimum of the effective potential, so small displacements experience a restoring force, and the mechanical frequency remains real. For the perturbative operating regime considered here, a stronger and experimentally more useful requirement is
\begin{equation}
 |G_F|\ll k_{m,0}.
\label{eq:cnt_stability}
\end{equation}
A positive $G_F$ reduces the effective CNT stiffness and therefore softens the mode, whereas a sufficiently large positive gradient drives the system toward static instability. A separate AFM stability requirement is that the attractive tip--CNT force gradient remains smaller than the effective cantilever stiffness $k_c$, thereby avoiding jump-to-contact \cite{Giessibl2003}. We note that the CNT condition in Equation \eqref{eq:cnt_stability} can be more restrictive because the effective stiffness of a suspended CNT may be much smaller than that of the AFM cantilever.

Dispersion forces provide an additional static interaction. As a simple order-of-magnitude benchmark, the Hamaker result for a spherical tip of radius $R$ above a planar surface gives \cite{Giessibl2003,Chang2026}
\begin{equation}
\begin{aligned}
|F_{\mathrm{vdW}}|&\simeq\frac{A_HR}{6z_0^2},\\
\left|\frac{\partial F_{\mathrm{vdW}}}{\partial z}\right|
&\simeq\frac{A_HR}{3z_0^3}.
\end{aligned}
\label{eq:vdw}
\end{equation}
where $A_H$ is the Hamaker constant and $z_0$ is the static tip--CNT separation. For $A_H=10^{-19}$ J, $R=20$ nm, and $z_0=50$ nm, these expressions give approximately $1.3\times10^{-13}$ N and $5.3\times10^{-6}$ N/m, respectively. These numbers should be regarded only as an approximate scale for the present geometry because a CNT is not a planar surface, and finite-geometry and retardation corrections can become relevant at separations of tens of nanometers \cite{Giessibl2003}. Nevertheless, the estimate illustrates that the static interaction can be orders of magnitude larger than the intended sub-fN resonant control force. Therefore, the relevant quantity for coherent actuation remains the small time-dependent force component after the static equilibrium and frequency shift have been calibrated.

\subsection{Tip-Position Noise and Frequency Noise}

Slow fluctuations of the AFM positioning system change the tip--CNT working separation and therefore modulate the force gradient. We denote these externally imposed position fluctuations by $\delta z(t)$, to distinguish them from the resonant CNT displacement $q(t)$. With the coordinate convention used in Section \ref{sec:driven}, the instantaneous gap can be written approximately as $z_{\mathrm{gap}}(t)=z_0+\delta z(t)-q(t)$.

Frequency fluctuations and dephasing are important in CNT mechanical resonators and in nanomechanical quantum control protocols \cite{Schneider2014,Remus2009,Chang2026}. Define the experimentally measurable tuning coefficient
\begin{equation}
 G_z
 \equiv
 \left.
 \frac{\partial\omega_m}{\partial z}
 \right|_{z_0},
\label{eq:Gz}
\end{equation}
where $G_z$ has units of angular frequency per unit displacement. To first order in the slow position fluctuation, we show that
\begin{equation}
 \delta\omega_m(t)=G_z\delta z(t).
\label{eq:freq_noise}
\end{equation}
If $\delta z$ is Gaussian with rms amplitude $\sigma_z$ and is effectively constant during one Ramsey sequence, it acts as quasi-static detuning noise \cite{Chang2026}. Averaging the accumulated phase over repeated realizations gives \cite{Schneider2014,Chang2026}
\begin{equation}
 \left\langle
 \ee^{-\ii\delta\omega_m t}
 \right\rangle
 =
 \exp\!\left[
 -\frac{1}{2}(G_z\sigma_z t)^2
 \right].
\label{eq:quasistatic_dephasing}
\end{equation}
If the Gaussian Ramsey envelope is written as $\exp[-(t/T_{\phi}^{*})^2]$, Equation \eqref{eq:quasistatic_dephasing} corresponds to $T_{\phi}^{*}=\sqrt{2}/(|G_z|\sigma_z)$.

For more general stationary fluctuations, it is useful to introduce a two-sided noise spectral density \cite{Clerk2010}. We define
\begin{equation}
 S_{zz}(\Omega)
 =
 \int_{-\infty}^{\infty}\dd t\,
 \ee^{\ii\Omega t}
 \left\langle
 \delta z(t)\delta z(0)
 \right\rangle ,
\label{eq:position_noise_psd}
\end{equation}
where $\Omega$ denotes the angular frequency at which the noise spectrum is evaluated. Equation \eqref{eq:freq_noise} then gives
\begin{equation}
 S_{\omega\omega}(\Omega)
 =
 G_z^2S_{zz}(\Omega).
\label{eq:freq_noise_psd}
\end{equation}
For a specified pulse sequence, this spectrum can be combined with the corresponding filter function to calculate the accumulated dephasing \cite{Clerk2010}.

For quasi-static noise, a useful conservative criterion is that the induced rms frequency fluctuation $\delta\omega_{\mathrm{rms}}=|G_z|\sigma_z$ remains small compared with both the intrinsic coherence linewidth and the anharmonic separation,
\begin{equation}
 \delta\omega_{\mathrm{rms}}
 \ll
 \Gamma_{2,\mathrm{int}},
\qquad
 \delta\omega_{\mathrm{rms}}
 \ll
 |K|.
\label{eq:freq_noise_conditions}
\end{equation}
Here $\Gamma_{2,\mathrm{int}}$ denotes the transverse decoherence rate of the lowest mechanical transition in the absence of AFM-induced frequency noise, and $K$ is the Kerr anharmonicity in angular-frequency units developed in Section \ref{sec:kerr}. The second condition is specific to spectrally selective operation of the anharmonic mode.

\begin{figure*}[t]

\centering
\includegraphics[width=0.94\textwidth]{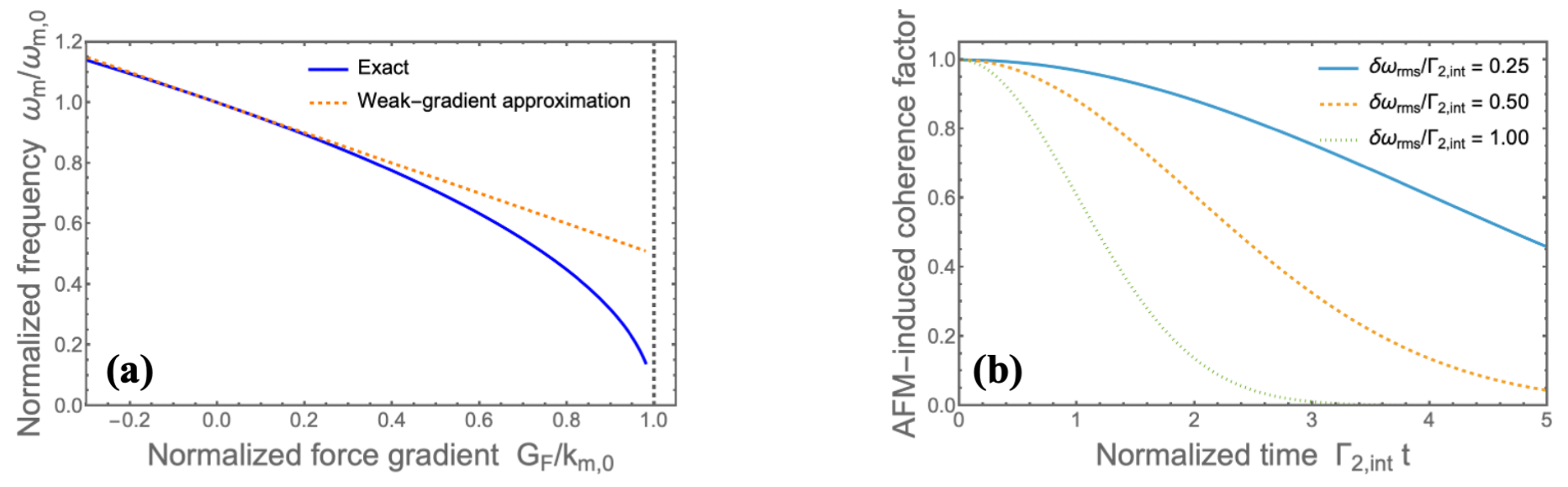}
\caption{Representative
 AFM--CNT backaction constraints.
(\textbf{a}) Normalized CNT resonance frequency $\omega_m/\omega_{m,0}$ as a function of the normalized static force gradient $G_F/k_{m,0}$, comparing the exact result (blue curve) with the weak-gradient approximation (dashed orange curve). The vertical dashed line marks the static-instability threshold $G_F/k_{m,0}=1$.
(\textbf{b}) AFM-induced coherence factor for quasi-static Gaussian tip-position noise at representative values of $\delta\omega_{\mathrm{rms}}/\Gamma_{2,\mathrm{int}}$, plotted as a function of normalized time $\Gamma_{2,\mathrm{int}}t$.}
\label{fig:afm_backaction}
\end{figure*}

The effects of the static force gradient and tip-position fluctuations are illustrated in Figure \ref{fig:afm_backaction}. Figure \ref{fig:afm_backaction}a shows the normalized mechanical resonance frequency as a function of the dimensionless force gradient $G_F/k_{m,0}$. The exact result from Equation \eqref{eq:gradient_frequency} is compared with the weak-gradient approximation of Equation \eqref{eq:frequency_shift_gradient}. The two expressions agree when $|G_F|\ll k_{m,0}$, while for a positive force gradient the exact resonance frequency decreases progressively as $G_F/k_{m,0}$ approaches unity, corresponding to the loss of static stability.

Figure \ref{fig:afm_backaction}b illustrates the corresponding effect of quasi-static tip-position fluctuations on mechanical coherence. The coherence factor from Equation \eqref{eq:quasistatic_dephasing} is plotted as a function of the normalized time $\Gamma_{2,\mathrm{int}}t$ for representative values of $\delta\omega_{\mathrm{rms}}/\Gamma_{2,\mathrm{int}}$, where $\delta\omega_{\mathrm{rms}}=|G_z|\sigma_z$. When the AFM-induced frequency fluctuations are small compared with the intrinsic coherence rate, the additional dephasing develops slowly. As $\delta\omega_{\mathrm{rms}}$ approaches $\Gamma_{2,\mathrm{int}}$, the coherence is suppressed on a time scale comparable to the intrinsic coherence time. These representative curves illustrate the motivation for the frequency-noise condition in Equation \eqref{eq:freq_noise_conditions}.

The equilibrium Brownian motion of the AFM cantilever provides a useful lower scale for the mechanical position noise. In the classical equipartition regime, $k_BT\gg\hbar\omega_c$, where $\omega_c$ is the cantilever angular frequency, the rms thermal displacement is \cite{Giessibl2003,Chang2026}
\begin{equation}
 x_{c,\mathrm{rms}}
 =
 \sqrt{\frac{k_BT}{k_c}},
\label{eq:cantilever_brownian}
\end{equation}
where $k_B$ is Boltzmann's constant, $T$ is the cantilever temperature, and $k_c$ is its effective stiffness. At $T=10$ mK, Equation \eqref{eq:cantilever_brownian} gives approximately $1.2$ pm for $k_c=0.1$ N/m, $0.37$ pm for $k_c=1$ N/m, and $0.12$ pm for $k_c=10$ N/m. These displacements are much smaller than a working separation of tens of nanometers. Instrumental vibration, drift, electrical cross-coupling, and low-frequency positioning noise can nevertheless exceed this equilibrium Brownian scale, so $S_{zz}(\Omega)$ must ultimately be characterized in the assembled cryogenic apparatus \cite{Clerk2010}.

\subsection{Voltage Noise, Resonant-Force Noise, and Heating}

Voltage fluctuations on the tip-bias line provide a second backaction channel. Under the same assumptions used in Section \ref{subsec:electrostatic_drive}, namely $V_{\mathrm{ac}}\ll V_{\mathrm{dc}}$ and small fluctuations $|\delta V|\ll|V_{\mathrm{dc}}|$, linearization of the electrostatic force gives \cite{Blencowe2004,Clerk2010,Chang2026}
\begin{equation}
 \delta F_V(t)
 \simeq
 \left|
 \frac{\partial C_t}{\partial z}
 \right|
 V_{\mathrm{dc}}\,
 \delta V(t),
\label{eq:voltage_force_noise}
\end{equation}
where $\delta V(t)$ is the fluctuating component of the tip--CNT voltage difference. The mode-shape normalization adopted in Section \ref{sec:driven} is assumed, so this local force is also the generalized force acting on the selected CNT mode. Using the same two-sided spectral convention as above, define \cite{Clerk2010}
\begin{equation}
 S_{FF}(\omega)
 =
 \int_{-\infty}^{\infty}\dd t\,
 \ee^{\ii\omega t}
 \left\langle
 \delta F(t)\delta F(0)
 \right\rangle.
\label{eq:force_psd_def}
\end{equation}
The voltage-induced contribution is then
\begin{equation}
 S_{FF}^{(V)}(\omega)
 =
 \left|
 \frac{\partial C_t}{\partial z}
 \right|^2
 V_{\mathrm{dc}}^2
 S_{VV}(\omega),
\label{eq:force_noise_psd}
\end{equation}
where $S_{VV}(\omega)$ is the corresponding two-sided voltage-noise spectral density.

To connect the force spectrum to mechanical heating, we write the quantized displacement as
\begin{equation}
 \hat q
 =
 x_{\mathrm{zpf}}
 \left(a+a^{\dagger}\right),
\qquad
 x_{\mathrm{zpf}}
 =
 \sqrt{\frac{\hbar}{2m_{\mathrm{eff}}\omega_m}},
\label{eq:zpf_backaction}
\end{equation}
where $a$ and $a^{\dagger}$ are the annihilation and creation operators of the mechanical mode and $x_{\mathrm{zpf}}$ is its zero-point displacement. Let $x_{01}=\langle0|\hat q|1\rangle$ denote the displacement matrix element between the two lowest vibrational states $|0\rangle$, $|1\rangle$ of the CNT. For weak anharmonicity, $x_{01}\simeq x_{\mathrm{zpf}}$ and $\omega_{01}\simeq\omega_m$.

With the spectral convention of Equation \eqref{eq:force_psd_def}, the upward transition rate generated by a general force-noise environment is \cite{Blencowe2004,Clerk2010}
\begin{equation}
 \Gamma_{\uparrow}^{\mathrm{AFM}}
 =
 \frac{x_{01}^2}{\hbar^2}
 S_{FF}^{\mathrm{AFM}}(-\omega_{01}).
\label{eq:heating_rate_general}
\end{equation}
For classical force noise the spectrum is symmetric, $S_{FF}(-\omega)=S_{FF}(\omega)$, so that in the weakly anharmonic limit this reduces to
\begin{equation}
 \Gamma_{\uparrow}^{\mathrm{AFM}}
 \simeq
 \frac{x_{\mathrm{zpf}}^2}{\hbar^2}
 S_{FF}^{\mathrm{AFM}}(\omega_m).
\label{eq:heating_rate}
\end{equation}
For a complete control-and-readout sequence of duration $t_{\mathrm{seq}}$, the short-time excitation probability is approximately $P_{\uparrow}\simeq\Gamma_{\uparrow}^{\mathrm{AFM}}t_{\mathrm{seq}}$. Negligible AFM-induced heating therefore requires \cite{Blencowe2004,Clerk2010}
\begin{equation}
 \Gamma_{\uparrow}^{\mathrm{AFM}}t_{\mathrm{seq}}\ll1.
\label{eq:heating_condition}
\end{equation}
This condition converts a measured voltage-noise or force-noise spectrum directly into an estimate of the excitation probability accumulated during one experimental sequence.

Low-frequency voltage fluctuations enter through a different channel. Because the static electrostatic force gradient scales approximately as $V_{\mathrm{dc}}^2$, slow fluctuations $\delta V(t)$ also modulate the effective spring constant and therefore $\omega_m$. Their contribution can be treated using the same frequency-noise framework as Equations \eqref{eq:freq_noise}--\eqref{eq:freq_noise_psd}. These two noise channels motivate careful cryogenic filtering and thermalization of the tip-bias line and favor the smallest static bias compatible with the required resonant force.

\subsection{Operating Conditions for Coherent Control and Backaction Suppression}

The preceding results can be combined into a compact operating window. For the effective two-level regime, a resonant force acting on the lowest mechanical transition produces the Rabi frequency \cite{RipsHartmann2013,Chang2026}
\begin{equation}
 \Omega_R
 =
 \frac{F_d x_{01}}{\hbar},
\label{eq:Rabi_force}
\end{equation}
where $x_{01}=\langle0|\hat q|1\rangle$ is the displacement matrix element, and $F_d$ is the resonant generalized force amplitude defined in Section \ref{subsec:electrostatic_drive}. A $\pi/2$ rotation of duration $t_{\pi/2}$ requires
\begin{equation}
 \Omega_R t_{\pi/2}=\frac{\pi}{2},
\qquad
 F_{\pi/2}
 =
 \frac{\pi\hbar}{2x_{01}t_{\pi/2}}.
\label{eq:Fpi2}
\end{equation}

The drive must be fast enough that coherence is preserved during the pulse, but sufficiently narrow in frequency that higher vibrational transitions are not appreciably excited. Let $\Delta\omega_p$ denote the characteristic spectral bandwidth of the control pulse. A useful set of operating conditions is

\begin{equation}
\begin{aligned}
&\Gamma_{2,\mathrm{int}}t_{\pi/2}\ll1,\\
&\max(\Omega_R,\Delta\omega_p)\ll|K|,\\
&|G_F|\ll k_{m,0},\\
&|G_z|\sigma_z\ll\min(\Gamma_{2,\mathrm{int}},|K|),\\
&\Gamma_{\uparrow}^{\mathrm{AFM}}t_{\mathrm{seq}}\ll1.
\end{aligned}
\label{eq:afm_window}
\end{equation}
The first two conditions constrain coherent and spectrally selective control. The remaining conditions require the static AFM force gradient, AFM-induced frequency fluctuations, and resonant-force noise heating to remain perturbative.

These requirements depend on experimentally distinct control parameters. For a fixed static working point, the coherent resonant force can be adjusted primarily through $V_{\mathrm{ac}}$, whereas the static electrostatic force gradient scales approximately as $V_{\mathrm{dc}}^2$ and depends strongly on the working separation $z_0$. Frequency noise additionally depends on the local tuning coefficient $G_z$ and on the mechanical stability of the AFM positioning system. This partial separation of control parameters provides an experimental means of increasing the resonant drive without proportionally increasing every backaction channel.

\section{Duffing/Kerr Dynamics and Active Cooling}
\label{sec:kerr}

\subsection{Quantization of the Duffing Mode}

We now quantize the same effective Duffing mode introduced in Section \ref{sec:driven}. In particular, $\omega_m$ denotes the small-amplitude mechanical frequency at the chosen static working point, including any static frequency shift produced by the AFM tip. Starting from Equation \eqref{eq:duffing_potential}, the conservative Hamiltonian is \cite{LifshitzCross2009,Semiao2009,RipsHartmann2013,Rips2014}
\begin{equation}
 H_m=
 \frac{\hat p^2}{2m_{\mathrm{eff}}}
 +\frac{1}{2}m_{\mathrm{eff}}\omega_m^2\hat q^2
 +\frac{\alpha}{4}\hat q^4.
\label{eq:quantum_duffing}
\end{equation}

The displacement operator is written in terms of the harmonic mode creation and annihilation operators as \cite{SchwabRoukes2005,Chang2026}
\begin{equation}
 \hat q=\zpf(a+a^{\dagger}),
 \qquad
 \zpf=
 \sqrt{\frac{\hbar}{2m_{\mathrm{eff}}\omega_m}}.
\label{eq:zpf}
\end{equation}
Thus, the zero-point displacement used here is defined using the same effective mechanical frequency $\omega_m$ that appears in the classical equation of motion.

For weak anharmonicity, the rapidly oscillating terms in $(a+a^{\dagger})^4$ can be neglected within the rotating-wave approximation (RWA). The number-conserving contribution is \cite{Semiao2009,Rips2014}
\begin{equation}
 (a+a^{\dagger})^4
 \longrightarrow
 6a^{\dagger 2}a^2+12a^{\dagger}a+3.
\label{eq:quartic_rwa}
\end{equation}
Substituting Equation \eqref{eq:quartic_rwa} into Equation \eqref{eq:quantum_duffing}, and absorbing the term linear in $a^{\dagger}a$ into the fundamental transition frequency, gives \cite{Semiao2009,RipsHartmann2013,Rips2014}
\begin{equation}
 H_m\simeq
 \hbar\omega_{01}a^{\dagger}a
 +\frac{\hbar K}{2}a^{\dagger 2}a^2,
\label{eq:kerr_hamiltonian}
\end{equation}
apart from an additive constant. With the quartic potential defined as $\alpha q^4/4$, the parameters in Equation \eqref{eq:kerr_hamiltonian} are
\begin{equation}
 \omega_{01}
 =
 \omega_m+\frac{3\alpha\zpf^4}{\hbar},
 \qquad
 K=
 \frac{3\alpha\zpf^4}{\hbar}.
\label{eq:K_relation}
\end{equation}
Here $\omega_{01}$ is the transition frequency between the two lowest vibrational states $|0\rangle$ and $|1\rangle$, while $K$ is the anharmonic separation in angular-frequency units. The adjacent transition frequencies are therefore
\begin{equation}
 \omega_{n,n+1}
 =
 \omega_{01}+nK,
 \qquad
 \omega_{12}-\omega_{01}=K.
\label{eq:transition_spacing}
\end{equation}
Equations \eqref{eq:quantum_duffing} and \eqref{eq:kerr_hamiltonian} define the relation between the Duffing coefficient $\alpha$ and the Kerr anharmonicity $K$ used throughout this paper. Equation \eqref{eq:K_relation} also makes the effect of mechanical softening transparent. For fixed $\alpha$ and $m_{\mathrm{eff}}$, $\zpf^4\propto\omega_m^{-2}$, so reducing the harmonic frequency increases the anharmonicity per phonon. More complete electromechanical models can also modify $\alpha$ itself, particularly close to a strong electrostatic softening regime. Nevertheless, electrostatic softening can enhance the effective nonlinearity sufficiently to support a nanomechanical two-level system (qubit) consistent with strongly nonlinear CNT behavior observed at low temperature \cite{RipsHartmann2013,Rips2014,Samanta2023}.

\subsection{Two-Level Selectivity and Drive-Induced Leakage}

The resonant AFM force introduced in Section \ref{sec:driven} couples linearly to the mechanical displacement according to \cite{RipsHartmann2013,Chang2026}
\begin{equation}
 H_d(t)=
 -F_d s(t)\cos(\omega_dt+\varphi_d)\hat q.
\label{eq:quantum_drive}
\end{equation}
When the mode is initialized near $|0\rangle$ and $\omega_d\simeq\omega_{01}$, a controlled two-level description requires the neighboring $|1\rangle\rightarrow|2\rangle$ transition to remain sufficiently far off resonance. Let $\Delta\omega_p$ denote the characteristic spectral bandwidth of the applied control pulse. Consistent with the operating conditions introduced in Section \ref{sec:backaction}, sufficient selectivity conditions are \cite{RipsHartmann2013,Rips2014,Chang2026}
\begin{equation}
 |K|\gg\Gamma_2,
 \qquad
 \Omega_R,\ \Delta\omega_p\ll|K|,
\label{eq:tls_conditions}
\end{equation}
where $\Gamma_2$ is the effective total transverse linewidth of the lowest transition, including intrinsic decoherence and AFM-induced broadening, and $\Omega_R$ is the resonant Rabi frequency defined in Equation \eqref{eq:Rabi_force}.   The neighboring $|1\rangle\rightarrow|2\rangle$ transition is detuned from the control frequency by approximately $K$. Perturbation theory therefore gives the leading leakage scaling \cite{Motzoi2009,Chang2026}
\begin{equation}
 P_{\mathrm{leak}}
\sim
 \frac{\Omega_R^2}{K^2}
\label{eq:leak_scaling}
\end{equation}
where the numerical coefficient depends on the pulse shape and on the matrix element of the neighboring transition. Smooth pulse shaping, including derivative-corrected protocols, can further suppress leakage \cite{Motzoi2009}.

A representative case from the nanomechanical-qubit literature is a CNT mode electrostatically softened to approximately $26.6$ MHz, for which the predicted separation between the $|0\rangle\rightarrow|1\rangle$ and $|1\rangle\rightarrow|2\rangle$ transitions is approximately $2.7$ MHz \cite{RipsHartmann2013,Chang2026}. The magnitude of this anharmonic separation depends on the specific CNT geometry and operating conditions, particularly the degree of electrostatic softening. The reported value nevertheless illustrates that MHz-scale transition separations are achievable in appropriately engineered CNT resonators. Because stronger softening also lowers the mechanical frequency and thereby increases the equilibrium thermal occupation, spectral selectivity and active cooling must be considered simultaneously.

\subsection{Thermal Occupation of Representative CNT Modes}

For a mechanical mode in thermal equilibrium at temperature $T$, the mean phonon occupation is given by the Bose--Einstein distribution \cite{SchwabRoukes2005,Aspelmeyer2014}
\begin{equation}
 \nth=
 \frac{1}
 {\exp(\hbar\omega_m/k_BT)-1}.
\label{eq:nth}
\end{equation}
At $T=10$ mK, representative CNT mechanical frequencies of $5.37$ GHz, $221$ MHz, and $54$ MHz correspond to thermal occupations of $\nth\simeq6.4\times10^{-12}$, $0.53$, and $3.38$, respectively. Thus, the GHz-frequency mode is already very close to its mechanical ground state under ideal thermal equilibrium conditions, whereas active cooling is required for the lower-frequency examples if the target occupation is substantially below one phonon.

Table \ref{tab:parameters} summarizes the thermal occupations together with representative zero-point amplitudes, $\pi/2$ force scales, relaxation times, and cooling requirements for the three CNT modes \cite{Chang2026}. The listed $F_{\pi/2}$ values are calculated using $t_{\pi/2}=100$ ns as a common reference for comparing the required control-force scales across the representative CNT modes. Longer pulses reduce the required force in inverse proportion to $t_{\pi/2}$ and may be needed to satisfy Equation \eqref{eq:tls_conditions} when $|K|$ is only a few megahertz. The final column gives the approximate cooling cooperativity required to reach $\bar n_f<0.1$ in the ideal resolved-sideband limit discussed below. The tabulated quantities correspond to the stated mechanical frequencies. If electrostatic softening by the AFM or another element shifts a device substantially from these frequencies, the frequency-dependent quantities, in particular $\zpf$ and $\nth$, must be recalculated for the actual operating point.

\begin{table*}[t]
\centering
\small
\setlength{\tabcolsep}{9pt}
\renewcommand{\arraystretch}{1.2}

\begin{tabular}{ccccccc}
\hline
$L$ (nm) &
$\omega_m/2\pi$ &
$\zpf$ (pm) &
$F_{\pi/2}$ (fN) &
$T_1$ ($\mu$s) &
$\nth$ &
$\mathcal{C}_{\mathrm{min}}$ \\
\hline
100  & 5.37 GHz & 2.14 & 0.77 & 0.29 & $6.4\times10^{-12}$ & $0$ \\
500  & 221 MHz  & 4.79 & 0.35 & 7.4  & 0.53 & $>4.3$ \\
1000 & 54 MHz   & 6.78 & 0.24 & 29.6 & 3.38 & $>32.8$ \\
\hline
\end{tabular}
\caption{Representative CNT parameters at $T=10$ mK. The frequencies,
zero-point amplitudes, $F_{\pi/2}$ values, and $T_1$ values correspond
to the representative geometries analyzed in Ref.~\cite{Chang2026},
with $Q_m=10^4$ and $F_{\pi/2}$ evaluated for $t_{\pi/2}=100$ ns.
The cooling cooperativity in the final column is defined in the next
subsection and is estimated using
$\bar n_f\simeq\nth/(1+\mathcal{C})$, assuming $n_{\min}\ll0.1$.}
\label{tab:parameters}
\end{table*}

\subsection{Resolved-Sideband Cooling Estimate}
\label{subsec:cooling}

Resolved-sideband cooling is relevant for the lower frequency and electrostatically softened CNT modes because their thermal phonon occupation can remain appreciable even at dilution refrigerator temperatures. Preparation with $\bar n_f\ll1$ provides the low-occupation initial state required for state-selective AFM control and phonon-sensitive CNT to CPB dispersive readout. The analysis presented below estimates the cooling strength required from an auxiliary driven electromagnetic or electromechanical resonator and does not rely on an optical cavity implementation. Measurement-based feedback and electronic transport mechanisms provide alternative routes to cooling when sufficiently strong electromechanical coupling is available \cite{Rossi2018,Santandrea2011,Stadler2016}. A standard route to active cooling is to couple the CNT to a driven electromagnetic resonator or another auxiliary degree of freedom that provides dynamical backaction \cite{WilsonRae2007,Teufel2011,Aspelmeyer2014}. To distinguish this cooling drive from the AFM control frequency $\omega_d$, let $\omega_p$ denote the electromagnetic pump frequency and $\omega_c$ the cavity resonance frequency. We define the cavity-pump detuning as
\begin{equation}
 \Delta_c=\omega_p-\omega_c.
\label{eq:cooling_detuning}
\end{equation}
Let $G$ denote the linearized cavity--mechanical coupling rate and $\kappa$ the cavity energy-decay linewidth, both expressed in angular-frequency units. Within the standard linearized weak-coupling treatment, the anti-Stokes and Stokes scattering rates are \cite{WilsonRae2007,Aspelmeyer2014}
\begin{align}
 A_-&=
 \frac{G^2\kappa}
 {(\kappa/2)^2+(\Delta_c+\omega_m)^2},
 \\
 A_+&=
 \frac{G^2\kappa}
 {(\kappa/2)^2+(\Delta_c-\omega_m)^2}.
\label{eq:sideband_rates}
\end{align}
With the convention $\Delta_c=\omega_p-\omega_c$, the anti-Stokes process $A_-$ is resonantly enhanced for red-sideband pumping near $\Delta_c=-\omega_m$. The resulting cooling-induced mechanical damping is \cite{WilsonRae2007,Aspelmeyer2014}
\begin{equation}
 \Gamma_{\mathrm{cool}}
 =
 A_--A_+.
\label{eq:cooling_rate}
\end{equation}
For red-sideband pumping with $\Delta_c\simeq-\omega_m$ and in the resolved-sideband regime $\kappa\ll\omega_m$, the standard approximations are \cite{WilsonRae2007,Teufel2011,Aspelmeyer2014}
\begin{equation}
 \Gamma_{\mathrm{cool}}
 \simeq
 \frac{4G^2}{\kappa},
 \qquad
 n_{\min}
 \simeq
 \left(
 \frac{\kappa}{4\omega_m}
 \right)^2.
\label{eq:resolved_sideband}
\end{equation}
Here $n_{\min}$ is the minimum phonon occupation imposed by finite sideband resolution in the idealized cooling limit. Combining the intrinsic mechanical bath with the cooling bath gives \cite{WilsonRae2007,Aspelmeyer2014}
\begin{equation}
 \bar n_f
 \simeq
 \frac{
 \gamma_m\nth+
 \Gamma_{\mathrm{cool}}n_{\min}
 }
 {
 \gamma_m+\Gamma_{\mathrm{cool}}
 }.
\label{eq:final_occupation}
\end{equation}
The damping rate $\gamma_m$ is the same intrinsic mechanical energy-decay rate used in Sections \ref{sec:driven} and \ref{sec:backaction}. It is convenient to define the cooling cooperativity \cite{Aspelmeyer2014}
\begin{equation}
 \mathcal{C}
 =
 \frac{\Gamma_{\mathrm{cool}}}{\gamma_m}
 \simeq
 \frac{4G^2}{\kappa\gamma_m}.
\label{eq:cooling_cooperativity}
\end{equation}
When $n_{\min}$ is negligible compared with a desired final occupation $\bar n_{\mathrm{tar}}$, Equations \eqref{eq:final_occupation} and \eqref{eq:cooling_cooperativity}, together with the condition $\bar n_f<\bar n_{\mathrm{tar}}$, give an approximate lower bound on the required cooperativity:
\begin{equation}
 \mathcal{C}>
 \frac{\nth}{\bar n_{\mathrm{tar}}}-1.
\label{eq:Crequirement}
\end{equation}
For the representative calculations below, we choose $\bar n_{\mathrm{tar}}=0.1$, corresponding to a thermal ground state population greater than approximately $90\%$. This provides a convenient criterion for comparing the cooling requirements of the different CNT modes.

\begin{figure*}[t]

\centering
\includegraphics[width=0.94\textwidth]{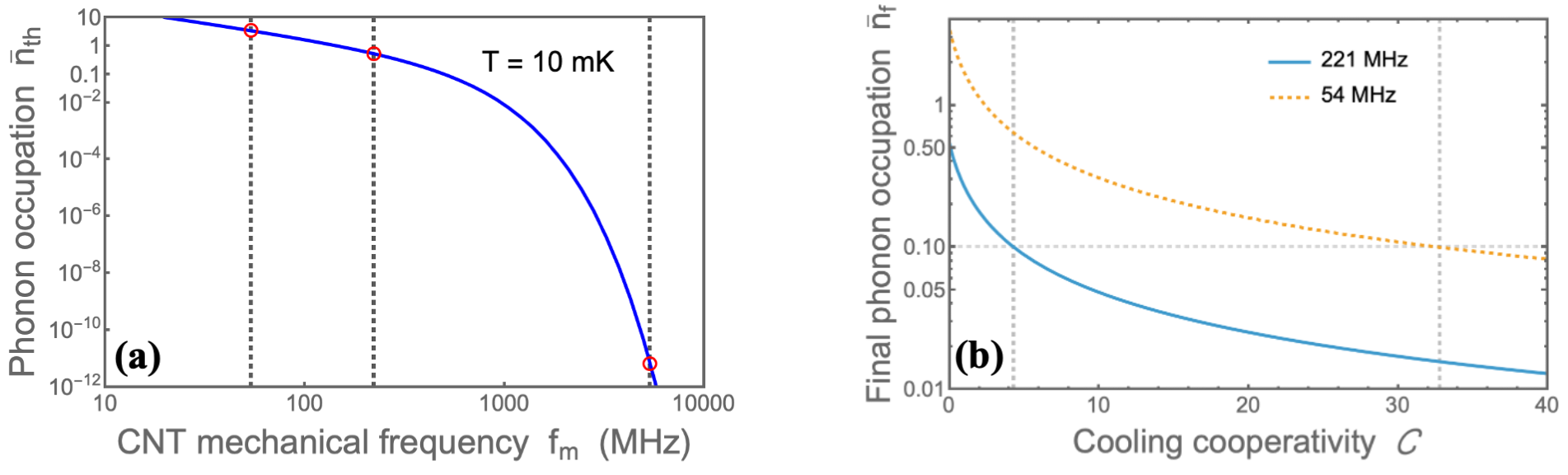}
\caption{Thermal
 occupation and active cooling of representative CNT modes.
(\textbf{a}) Thermal phonon occupation $\nth$ at $T=10$ mK as a function of CNT mechanical frequency. Red circles indicate the representative modes at $54$ MHz, $221$ MHz, and $5.37$ GHz.
(\textbf{b}) Final phonon occupation $\bar n_f$ as a function of cooling cooperativity $\mathcal{C}$ for the $221$ MHz mode (blue continuous curve) and the $54$ MHz mode (dotted orange curve), using $\bar n_f\simeq\nth/(1+\mathcal{C})$.}
\label{fig:cooling}
\end{figure*}

The dependence of thermal occupation and cooling performance on the mechanical frequency is illustrated in Figure \ref{fig:cooling}. Figure \ref{fig:cooling}a shows the equilibrium thermal phonon occupation at $T=10$ mK as a function of CNT mechanical frequency, with the representative modes at $54$ MHz, $221$ MHz, and $5.37$ GHz indicated by red circles. The strong decrease of $\nth$ with increasing frequency shows why the GHz mode is already close to its ground state at base temperature, whereas the lower-frequency modes benefit from active cooling. Figure \ref{fig:cooling}b shows the corresponding final phonon occupation for the $221$ MHz and $54$ MHz modes as a function of cooling cooperativity, using the approximation $\bar n_f\simeq\nth/(1+\mathcal{C})$. Consistent with the cooling requirements summarized in Table \ref{tab:parameters}, reaching $\bar n_f<0.1$ requires $\mathcal{C}>4.3$ for the $221$ MHz mode and $\mathcal{C}>32.8$ for the $54$ MHz mode. Thus, operation at lower mechanical frequency imposes a substantially greater cooling requirement.

Equations \eqref{eq:sideband_rates}--\eqref{eq:Crequirement} represent linearized harmonic mode estimates for the initialization stage. If the Kerr separation becomes comparable to the cavity linewidth or to the cooling bandwidth, the individual mechanical transitions become spectrally distinguishable and a transition-resolved treatment of cooling is more appropriate \cite{Rips2014}. Measurement-based feedback provides an alternative when a sufficiently efficient continuous displacement measurement is available \cite{Rossi2018}. In that case, the achievable occupation depends on measurement backaction, feedback bandwidth, and detection efficiency rather than on the sideband limit of Equation \eqref{eq:resolved_sideband}. For the present architecture, a practical strategy is to use an on-chip electromechanical or cavity channel for initialization and cooling, the AFM primarily for localized coherent actuation, and the CPB for state-sensitive readout after the control or parity-mapping interval \cite{Chang2026}.

\section{CNT--CPB Coupling and Dispersive Readout}
\label{sec:cpb}

The analysis in this section builds on our previous treatment of AFM-driven coherent control and quantum tomography of a suspended CNT \cite{Chang2026}, together with established models of capacitive coupling between nanomechanical resonators and Cooper-pair boxes (CPBs) \cite{Armour2002, Persson2010,Blencowe2004,Pirkkalainen2013,Lee2023, Suh2010}. In these systems, mechanical displacement modulates a CPB gate capacitance and therefore its dimensionless gate charge. Near the CPB charge-degeneracy point, this interaction becomes predominantly transverse in the qubit energy basis, leading to a Jaynes--Cummings-type interaction. In the large-detuning regime, the same coupling produces a phonon number-dependent dispersive shift that can be converted into a CPB population by a Ramsey sequence \cite{Blais2004,Zueco2009,Blais2021}. We use this mechanism to connect the AFM-controlled CNT dynamics developed above to a state-sensitive superconducting readout.

\subsection{Displacement-Dependent Gate Charge}

A CPB consists of a superconducting island connected to a reservoir by a Josephson junction with Josephson energy $E_J$ and junction capacitance $C_J$ \cite{Armour2002, Persson2010}. The total island capacitance is denoted by $C_\Sigma$, and the single-electron charging energy is $E_C=e^2/(2C_\Sigma)$. It is convenient to define $E_Q=4E_C=(2e)^2/(2C_\Sigma)$, where $e$ is the elementary charge.

We use the same generalized CNT displacement coordinate $q$ introduced in Section \ref{sec:driven}. Let $C_m(q)$ denote the capacitance between the CNT and the CPB island and $V_m$ the corresponding bias voltage. Additional static gates are represented by a capacitance $C_g$ and bias $V_g$. The dimensionless CPB gate charge is then \cite{Armour2002,Blencowe2004,Persson2010}
\begin{equation}
 n_g(q)=
 \frac{C_gV_g+C_m(q)V_m+\cdots}{2e}.
\label{eq:ngq}
\end{equation}
For small motion about the static operating point,
\begin{equation}
 C_m(q)
 \simeq
 C_m^{(0)}+C_m' q,
\qquad
 C_m'=
 \left.
 \frac{\partial C_m}{\partial q}
 \right|_{q=0},
\label{eq:Cm_linear}
\end{equation}
so that
\begin{equation}
 n_g(q)=n_{g0}+\eta_g q,
\qquad
 \eta_g=\frac{V_m C_m'}{2e},
\label{eq:ng_linear}
\end{equation}
where $n_{g0}=n_g(0)$ is the static gate charge. Because $q$ is the generalized coordinate used throughout the manuscript, $C_m'$ includes the appropriate mode-shape factor if the CPB gate is located away from the AFM actuation point (see Section \ref{sec:driven}).

Near a charge-degeneracy point, only two CPB charge states need to be retained. In this charge basis, the CPB Hamiltonian can be written \cite{Blencowe2004,Persson2010}
\begin{equation}
 H_{\mathrm{CPB}}
 =
 -\frac{1}{2}\epsilon(n_g)\tau_z
 -\frac{E_J}{2}\tau_x,
\qquad
 \epsilon(n_g)=E_Q(1-2n_g),
\label{eq:cpb_charge_hamiltonian}
\end{equation}
where $\tau_x$ and $\tau_z$ are Pauli operators in the charge basis. Substituting Equation \eqref{eq:ng_linear} into Equation \eqref{eq:cpb_charge_hamiltonian} gives the displacement-dependent interaction
\begin{equation}
 H_{\mathrm{int}}
 =
 E_Q\eta_g\hat q\,\tau_z
 =
 \hbar\lambda(a+a^\dagger)\tau_z,
\label{eq:longitudinal_charge_coupling}
\end{equation}
with
\begin{equation}
 \lambda
 =
 \frac{E_Q\eta_g\zpf}{\hbar}
 =
 \frac{4E_C}{\hbar}
 \left(
 \frac{V_m C_m'}{2e}
 \right)\zpf.
\label{eq:lambda}
\end{equation}
Here $\lambda$ is the CNT--CPB coupling rate in the charge basis, and $\zpf$ is the zero-point amplitude defined in Equation \eqref{eq:zpf}.

\subsection{Energy Eigenbasis and Jaynes--Cummings Interaction}

At the static operating point, define
\begin{equation}
 \epsilon_0=E_Q(1-2n_{g0}),
\qquad
 \hbar\omega_q=
 \sqrt{\epsilon_0^2+E_J^2},
\label{eq:qubit_gap}
\end{equation}
where $\omega_q$ is the CPB transition angular frequency. Let $\sigma_i$ denote Pauli operators in the CPB energy eigenbasis, with $|g\rangle$ and $|e\rangle$ denoting the CPB ground and excited energy eigenstates, respectively, so that $\sigma_z=|e\rangle\langle e|-|g\rangle\langle g|$. Introducing the mixing angle $\theta$ through
\begin{equation}
 \cos\theta=
 \frac{\epsilon_0}{\hbar\omega_q},
\qquad
 \sin\theta=
 \frac{E_J}{\hbar\omega_q},
\label{eq:mixing_angle}
\end{equation}
the charge operator becomes
\begin{equation}
 \tau_z=
 -\cos\theta\,\sigma_z
 +\sin\theta\,\sigma_x.
\label{eq:charge_to_energy}
\end{equation}
The sign of the longitudinal term follows from the eigenstate convention used above and does not affect the magnitude of the transverse coupling. Using the mechanical Hamiltonian $H_m$ from Equation \eqref{eq:kerr_hamiltonian}, the coupled system is therefore \cite{Armour2002,Blencowe2004,Pirkkalainen2013}
\begin{equation}
\begin{aligned}
H&=H_m+\frac{\hbar\omega_q}{2}\sigma_z\\
&\quad+\hbar\lambda(a+a^\dagger)
(-\cos\theta\,\sigma_z+\sin\theta\,\sigma_x).
\end{aligned}
\label{eq:full_cpb_mech}
\end{equation}
At the charge-degeneracy point, $n_{g0}=1/2$, so $\epsilon_0=0$ and $\theta=\pi/2$. The interaction is then transverse. For weak coupling, $|g_m|\ll\omega_{01},\omega_q$, the rotating-wave approximation gives the Jaynes--Cummings-type interaction \cite{Armour2002,Blencowe2004,Blais2004}
\begin{equation}
 H_{\mathrm{JC}}
 \simeq
 H_m+
 \frac{\hbar\omega_q}{2}\sigma_z
 +\hbar g_m
 \left(
 a\sigma_+ +a^\dagger\sigma_-
 \right),
\label{eq:JC}
\end{equation}
where
\begin{equation}
 g_m=\lambda\sin\theta
\label{eq:gm}
\end{equation}
is the transverse CNT--CPB coupling rate and $\sigma_\pm=(\sigma_x\pm i\sigma_y)/2$ are the CPB raising and lowering operators.

\subsection{Dispersive Interaction}

Since the harmonic frequency $\omega_m$ differs from the lowest mechanical transition frequency $\omega_{01}$ because of the Kerr anharmonicity, as discussed in Section \ref{sec:kerr}, we define the CNT--CPB detuning with respect to the transition frequency,
\begin{equation}
\Delta_q=\omega_q-\omega_{01}.
\label{eq:Deltaq}
\end{equation}
For the appreciably occupied phonon numbers $n$, the dispersive approximation requires \cite{Blais2004,Zueco2009,Boissonneault2009,Blais2021}
\begin{equation}
 |\Delta_q|
 \gg
 |g_m|\sqrt{n+1}.
\label{eq:dispersive_condition}
\end{equation}
For the simple phonon-independent dispersive shift used below, the change in mechanical transition frequency produced by the Kerr term must also remain small compared to the qubit detuning over the relevant occupation range,
\begin{equation}
 |nK|\ll|\Delta_q|.
\label{eq:kerr_dispersive_condition}
\end{equation}
Within the dispersive regime defined above, a Schrieffer--Wolff transformation \cite{Blais2004,Zueco2009,Blais2021} with generator
\begin{equation}
 S=
 \frac{g_m}{\Delta_q}
 \left(
 a\sigma_+-a^\dagger\sigma_-
 \right)
\label{eq:SW_generator}
\end{equation}
gives, to second order in $g_m/\Delta_q$,
\begin{equation}
\begin{aligned}
H_{\mathrm{disp}}&\simeq H_m
+\frac{\hbar\widetilde\omega_q}{2}\sigma_z\\
&\quad+\frac{\hbar\chi}{2}a^\dagger a\,\sigma_z.
\end{aligned}
\label{eq:Hdisp}
\end{equation}
where
\begin{equation}
 \widetilde{\omega}_q
 =
 \omega_q+\frac{g_m^2}{\Delta_q}
\end{equation}
is the CPB transition frequency including the phonon-independent dispersive Lamb shift, and
\begin{equation}
 \chi=
 \frac{2g_m^2}{\Delta_q}.
\label{eq:chi}
\end{equation}
With this convention, a mechanical Fock state containing $n$ phonons changes the CPB transition frequency by $\chi n$. Because both $H_m$ and the dispersive interaction are functions of $\hat n=a^\dagger a$, the Hamiltonian in Equation \eqref{eq:Hdisp} commutes with $\hat n$ and therefore permits a quantum nondemolition measurement of phonon number to this order. If the Kerr-induced variation of the individual mechanical transition frequencies becomes comparable to $\Delta_q$, the dispersive shift becomes phonon number dependent, and the simple constant-$\chi$ approximation must be replaced by a transition-resolved treatment \cite{Rips2014,Zueco2009,Boissonneault2009,Blais2021}.

\subsection{Ramsey Phase Mapping and Phonon Parity}

Ramsey interferometry was used in our previous analysis as a phase-sensitive probe of the AFM-driven CNT resonator \cite{Chang2026}. Starting from the mechanical ground state, a calibrated AFM $\pi/2$ pulse prepares the CNT in a superposition $(|0\rangle+|1\rangle)/\sqrt{2}$. After a free evolution interval $\tau$, a second AFM $\pi/2$ pulse converts the accumulated phase into a measurable population. For a mechanical-drive detuning $\Delta_d=\omega_d-\omega_{01}$, the Ramsey phase evolves as $\Delta_d\tau$, while decay of the fringe contrast provides the mechanical transverse coherence time $T_{2,m}$. The CPB protocol used here employs the same interferometric principle, but the Ramsey superposition is prepared in the CPB and the accumulated phase is generated by the phonon number-dependent dispersive interaction.

The sequence begins with a CPB $\pi/2$ pulse that prepares the state $(|g\rangle+|e\rangle)/\sqrt{2}$ with a first CPB $\pi/2$ pulse. During a dispersive interaction interval $t$, Equation \eqref{eq:Hdisp} produces the evolution \cite{Blencowe2004,Blais2004,Chang2026}
\begin{equation}
 U_{\mathrm{int}}(t)
 =
 \exp\!\left[
 -i\frac{\chi t}{2}
 \hat n\,\sigma_z
 \right].
\label{eq:Uint}
\end{equation}
For a mechanical Fock component $|n\rangle$, the relative phase between the two CPB states is therefore
\begin{equation}
 \varphi_n(t)=\chi nt.
\label{eq:number_phase}
\end{equation}
Choosing the parity-mapping interval
\begin{equation}
 t_{\pi}=
 \frac{\pi}{|\chi|}
\label{eq:tpi_parity}
\end{equation}
gives a relative phase $\pi n$, up to the sign of $\chi$, and hence the parity factor
\begin{equation}
 \Pi=(-1)^{\hat n}=\exp(i\pi\hat n).
\label{eq:parity_operator}
\end{equation}
Here $t_{\pi}$ denotes the dispersive parity-mapping interval and is distinct from the AFM $\pi/2$ pulse duration $t_{\pi/2}$ introduced previously.

A second CPB rotation with opposite phase converts the number-dependent phase into population. For the pulse convention $\{R_{\pi/2},t_\pi,R_{-\pi/2}\}$, the probability of the CPB excited state is:
\cite{Chang2026}
\begin{equation}
 P_e=
 \frac{1}{2}
 \left[
 1-\langle\Pi\rangle
 \right].
\label{eq:Pe_parity}
\end{equation}
If the AFM first applies the phase-space displacement
\begin{equation}
 D(\alpha)=
 \exp\!\left(
 \alpha a^\dagger-\alpha^*a
 \right),
\label{eq:Dalpha_cpb}
\end{equation}
where $\alpha$ is the complex displacement amplitude, the resulting displaced-parity measurement gives the CNT Wigner function \cite{Chang2026}
\begin{equation}
 W(\alpha)=
 \frac{2}{\pi}
 \mathrm{Tr}
 \left[
 D(-\alpha)\rho_mD(\alpha)\Pi
 \right],
\label{eq:Wigner_cpb}
\end{equation}
where $\rho_m$ is the mechanical density operator. With the pulse convention of Equation \eqref{eq:Pe_parity},
\begin{equation}
 P_e(\alpha)=
 \frac{1}{2}
 \left[
 1-\frac{\pi}{2}W(\alpha)
 \right].
\label{eq:Pe_Wigner}
\end{equation}

High-contrast parity mapping requires the dispersive interaction to act for many mechanical carrier cycles while remaining short compared with the relevant CPB and mechanical relaxation times,
\begin{equation}
 \frac{2\pi}{\omega_{01}}
 \ll
 t_{\pi}
 \ll
 T_{2,q},\,T_{1,m},
\label{eq:parity_times}
\end{equation}
where $T_{2,q}$ is the CPB coherence time during the mapping interval and $T_{1,m}$ is the mechanical energy-relaxation time. For protocols that require preservation of mechanical phase coherence throughout the full preparation and measurement sequence, the total sequence duration must also remain short compared with $T_{2,m}$. In terms of $\chi$, Equation \eqref{eq:parity_times} gives the approximate conditions
\begin{equation}
\begin{aligned}
\frac{1}{2T_{2,q}}&\ll\frac{|\chi|}{2\pi},\\
\frac{1}{2T_{1,m}}&\ll\frac{|\chi|}{2\pi}\ll\frac{f_{01}}{2},\\
f_{01}&=\frac{\omega_{01}}{2\pi}.
\end{aligned}
\label{eq:chi_window}
\end{equation}

\subsection{Quantum-Capacitance Readout of the CPB}
\label{subsec:quantum_capacitance}

After the CNT phonon information has been mapped onto the CPB, the CPB state can be detected electrically through its quantum capacitance. This method uses the state-dependent curvature of the CPB energy bands with respect to gate charge \cite{Duty2005,Sillanpaa2005,Persson2010}. From the Hamiltonian given in Equation \eqref{eq:cpb_charge_hamiltonian}, the two CPB eigenenergies are
\begin{equation}
 E_{g/e}
 =
 \mp\frac{\Delta E}{2},
\qquad
 \Delta E=
 \sqrt{
 E_J^2+E_Q^2(1-2n_g)^2
 }.
\label{eq:cpb_eigenenergies}
\end{equation}
For a gate capacitance $C_g$ used for rf readout, we define the state-dependent quantum capacitance by \cite{Duty2005,Sillanpaa2005,Persson2010}
\begin{equation}
 C_Q^{(s)}
 =
 \left(
 \frac{C_g}{2e}
 \right)^2
 \frac{\partial^2E_s}{\partial n_g^2},
\qquad
 s=g,e.
\label{eq:CQ_definition}
\end{equation}
Using Equation \eqref{eq:cpb_eigenenergies} in this definition yields
\begin{equation}
 C_Q^{(g/e)}
 =
 \mp
 \frac{C_g^2}{C_\Sigma}
 \frac{E_QE_J^2}{\Delta E^3}.
\label{eq:CQ}
\end{equation}
Thus, the ground and excited states have quantum capacitances of equal magnitude and opposite sign. At the charge-degeneracy point, $n_g=1/2$ and $\Delta E=E_J$, so
\begin{equation}
 C_Q^{(g/e)}(1/2)
 =
 \mp
 \frac{C_g^2}{C_\Sigma}
 \frac{E_Q}{E_J}.
\label{eq:CQ_degeneracy}
\end{equation}
The charge-degeneracy point therefore remains insensitive to first-order gate-charge fluctuations while retaining a state-dependent energy curvature that can be used for readout.

Consider the readout gate as part of an $LC$ resonator with inductance $L_r$, baseline effective capacitance $C_r$, resonance frequency $\omega_r\simeq1/\sqrt{L_rC_r}$, and quality factor $Q_r$. A small state-dependent change in quantum capacitance, $|\delta C_Q|\ll C_r$, shifts the resonator frequency according to \cite{Persson2010}
\begin{equation}
 \frac{\delta\omega_r}{\omega_r}
 \simeq
 -\frac{\delta C_Q}{2C_r}.
\label{eq:resonator_shift}
\end{equation}
Defining
$\Delta C_Q=C_Q^{(e)}-C_Q^{(g)}$, the magnitude of the reflected-phase contrast near resonance is approximately \cite{Persson2010}
\begin{equation}
 |\Delta\phi|
 \simeq
 2\arctan\!\left(
 Q_r\frac{|\Delta C_Q|}{C_r}
 \right),
\label{eq:phase_contrast}
\end{equation}
which reduces to the expression $|\Delta\phi|\simeq2Q_r|\Delta C_Q|/C_r$ for small phase contrast.

Quantum-capacitance measurements of superconducting charge devices have demonstrated state discrimination on time scales of the order of tens of nanoseconds \cite{Duty2005,Sillanpaa2005,Persson2010}. In particular, Persson and collaborators obtained a signal-to-noise ratio of order unity for $50$--$100$ ns readout pulses using resonators with $Q_r$ of order $10^2$, with the shortest measurement time limited by a resonator time constant of approximately $60$ ns \cite{Persson2010}. Their experiment used quasiparticle parity to characterize the CPB readout. By contrast, in the present architecture, the CPB serves as an intermediate quantum sensor for the CNT: mechanical phonon parity is first mapped coherently onto the CPB state through the dispersive interaction, and the quantum-capacitance measurement is then used only for the final electrical discrimination of that mapped state.

\subsection{Readout Backaction and Quasiparticle Considerations}

The dispersive condition in Equation \eqref{eq:dispersive_condition} suppresses real CNT--CPB energy exchange, but it does not eliminate readout backaction. Fluctuations in the gate voltage and rf photons can dephase or relax the CPB, while CPB-state fluctuations during the dispersive interval can modulate the CNT frequency through Equation \eqref{eq:Hdisp}. Nonequilibrium quasiparticles provide an additional source of switching and charge-parity fluctuations \cite{Aumentado2004,Persson2010}.

These considerations favor a staged measurement sequence in which the readout resonator is weakly populated during AFM control and dispersive CNT--CPB mapping and is driven strongly only after the second CPB Ramsey rotation. If $t_{\mathrm{ro}}$ denotes the final electrical readout time and $T_{1,q}$ the CPB energy-relaxation time, a basic requirement is
\begin{equation}
 t_{\mathrm{ro}}\ll T_{1,q}.
\label{eq:readout_time_condition}
\end{equation}
The total measurement cycle is therefore determined jointly by the coherent mapping time $t_\pi$ and the subsequent CPB readout time. Quasiparticle parity of the superconducting island must be monitored independently because it is distinct from the CNT phonon parity that constitutes the intended mechanical observable.

\begin{table*}[t]
\centering
\small
\setlength{\tabcolsep}{8pt}
\renewcommand{\arraystretch}{1.5}

\begin{tabular}{@{}ll@{}}
\hline
Physical requirement & Operating conditions \\
\hline

AFM stability and frequency noise &
$\begin{gathered}
|G_F|\ll k_{m,0}, \qquad
|G_z|\sigma_z\ll\Gamma_{2,\mathrm{int}},\qquad
|G_z|\sigma_z\ll|K|
\end{gathered}$ \\

AFM-induced heating &
$\Gamma_{\uparrow}^{\mathrm{AFM}}t_{\mathrm{seq}}\ll1$ \\

Mechanical-state selectivity &
$\begin{gathered}
\bar n_f\ll1,\qquad
\Gamma_2\ll|K|,\qquad
\Omega_R\ll|K|,\qquad
\Delta\omega_p\ll|K|
\end{gathered}$ \\

Dispersive CNT--CPB coupling &
$\begin{gathered}
|g_m|\sqrt{n_{\max}+1}\ll|\Delta_q|,\\
n_{\max}|K|\ll|\Delta_q|
\end{gathered}$ \\

Phonon-parity mapping &
$\displaystyle
\frac{2\pi}{\omega_{01}}
\ll t_\pi=\frac{\pi}{|\chi|}
\ll T_{2,q},\,T_{1,m}$ \\

Final CPB readout &
$t_{\mathrm{ro}}\ll T_{1,q}$ \\

\hline
\end{tabular}
\caption{Summary of operating conditions for coherent AFM control,
nonlinear CNT dynamics, and dispersive CPB readout.}
\label{tab:operating_conditions}
\end{table*}

\section{Combined Operating Window and Experimental Sequence}
\label{sec:window}

The proposed experiments integrate three closely connected elements: localized AFM control of the CNT, nonlinear mechanical-state preparation and selectivity, and dispersive coupling to a CPB for superconducting readout. Their simultaneous implementation requires a compatible range of mechanical stability, coherence, coupling strengths, noise levels, and experimental time scales.

\subsection{Hierarchy of Scales}

The constraints derived for these different stages can be collected into a common hierarchy that defines the operating regime of the complete device \cite{Giessibl2003,Clerk2010,RipsHartmann2013,Rips2014,Blais2004,Blais2021,Chang2026}. In particular, coherent control must remain sufficiently strong and spectrally selective while AFM-induced backaction, thermal occupation, dispersive exchange, and readout errors remain small. A compact set of sufficient operating conditions is given in Table~\ref{tab:operating_conditions}. Here $k_{m,0}$ is the reference CNT stiffness before the local AFM force gradient is included, $\bar n_f$ is the final phonon occupation after cooling, $\Delta\omega_p$ is the spectral bandwidth of the mechanical control pulse, and $n_{\max}$ is the largest significantly occupied mechanical Fock level. The remaining quantities were defined in the preceding sections. For phase-sensitive protocols such as Wigner tomography, the preparation and mapping sequence should also remain short compared with the mechanical coherence time $T_{2,m}$.

These conditions expose two important trade-offs. Electrostatic softening can increase the Kerr separation $|K|$, but the accompanying reduction of the mechanical frequency increases the thermal occupation and therefore the cooling requirement. Similarly, increasing the CNT--CPB coupling $g_m$ increases the dispersive shift $|\chi|=2g_m^2/|\Delta_q|$ and shortens the parity-mapping interval, but stronger capacitive coupling can also increase sensitivity to electrical noise and backaction. The preferred operating regime therefore balances these effects rather than maximizing a single coupling parameter.

\subsection{Experimental Calibration Sequence}

A practical calibration procedure can be implemented in several stages. First, with the CPB readout inactive, the CNT resonance can be measured as a function of AFM tip--CNT separation $z_0$ and dc bias $V_{\mathrm{dc}}$ to determine the static force-gradient effects, the tuning coefficient $G_z$, and the range of stable electrostatic softening \cite{Giessibl2003,Chang2026}. Weak continuous-wave excitation can then be used to determine the linear response and $Q_m$, followed by measurements of the nonlinear amplitude and phase response. Fits to Equations \eqref{eq:duffing_response} and \eqref{eq:duffing_phase} determine the effective Duffing coefficient $\alpha$ and the nonlinear damping coefficient $\eta$ \cite{LifshitzCross2009,Eichler2011}.

The AFM drive can next be calibrated in the time domain using the transient response of Equation \eqref{eq:transient_green} and, in the two-level regime, mechanical Rabi oscillations \cite{RipsHartmann2013,Chang2026}. Measurements of the position-noise spectrum $S_{zz}(\Omega)$ and voltage-noise spectrum $S_{VV}(\omega)$ then provide direct estimates of AFM-induced dephasing and heating \cite{Clerk2010} through the relations developed in Section \ref{sec:backaction}.

The CNT is subsequently cooled, and its final mean phonon occupation $\bar n_f$ is determined. For lower frequency or strongly softened modes, the relations in Section \ref{subsec:cooling} provide target values for the cavity linewidth $\kappa$, linearized cavity--mechanical coupling $G$, and cooling cooperativity $\mathcal{C}$ \cite{WilsonRae2007,Teufel2011,Aspelmeyer2014}. The CPB can then be characterized independently of the CNT--CPB dispersive protocol by locating the charge-degeneracy point, measuring its transition frequency $\omega_q$, and calibrating the state-dependent quantum-capacitance response \cite{Duty2005,Sillanpaa2005,Persson2010}.

Finally, the CNT--CPB coupling can be used to determine $g_m$, $\Delta_q$, and the dispersive shift $\chi$. A CPB Ramsey measurement as a function of interaction time provides an experimental determination of the phonon number-dependent phase and the parity-mapping interval $t_{\pi}$ \cite{Blais2004,Chang2026}. This staged procedure allows each part of the architecture to be tested before the complete protocol is implemented. Deviations from the Duffing model, excessive AFM backaction, breakdown of the dispersive approximation, or insufficient CPB readout contrast can then be identified independently.

\section{Discussion}
\label{sec:discussion}

The principal result of this work is a unified quantitative framework for integrating localized AFM control, Duffing/Kerr nonlinear dynamics, active cooling, and superconducting dispersive readout in suspended-carbon nanotube quantum electromechanics. By expressing these processes in terms of experimentally measurable parameters, the analysis identifies the simultaneous conditions required for coherent control, limited backaction, state-selective dynamics, and rapid quantum readout. It extends our previous proposal for AFM-driven coherent control and quantum tomography of a CNT \cite{Chang2026} by explicitly incorporating actuation-induced backaction, the connection between classical Duffing response and the quantum Kerr spectrum, cooling requirements associated with electrostatic softening, and a complete CNT--CPB dispersive mapping and quantum-capacitance readout sequence.

An important consequence of the AFM analysis is that the magnitude of the resonant control force alone is not an adequate measure of backaction. The representative CNTs considered here require sub-femtonewton forces for coherent control, with approximately $0.24$--$0.77$ fN required for a $100$ ns $\pi/2$ pulse \cite{Chang2026}. These forces are much smaller than typical static AFM tip--sample interactions. Nevertheless, the different components of the AFM interaction affect the CNT through physically distinct channels. The resonant Fourier component provides coherent actuation, the static force shifts the equilibrium position, the force gradient renormalizes the mechanical stiffness, slow tip-position and voltage fluctuations produce frequency noise and dephasing, and resonant-force noise can directly excite the mechanical mode \cite{Giessibl2003,Clerk2010}. Consequently, a small coherent drive does not by itself imply negligible measurement disturbance. The useful operating regime is determined instead by the ratio of the desired resonant control rate to these separate backaction channels. This distinction is particularly relevant for scanning-probe control of quantum nanomechanical systems, where the AFM tip is necessarily positioned close to the object being manipulated.

The nonlinear analysis reveals a second design constraint. Electrostatic softening can increase the zero-point displacement and thereby enhance the Kerr separation associated with a given effective quartic nonlinearity. This favors spectral resolution of the lowest mechanical transitions and the realization of an effective nanomechanical two-level system \cite{RipsHartmann2013,Rips2014}. At the same time, reducing the mechanical frequency increases the equilibrium thermal occupation and places greater demands on active cooling. The representative calculations illustrate this competition. At $10$ mK, the $5.37$ GHz mode is essentially in its ground state under ideal thermalization, whereas the $221$ MHz and $54$ MHz modes have thermal occupations of approximately $0.53$ and $3.38$, respectively. Within the resolved-sideband estimate used here, cooling the latter two modes to $\bar n_f<0.1$ requires cooperativities greater than approximately $4.3$ and $32.8$. These values should not be interpreted as identifying a universal optimum frequency, since the achievable Kerr coefficient, coupling strengths, quality factor, and technical noise are device-dependent. They do suggest, however, that intermediate frequencies in the hundreds of megahertz range can provide an attractive compromise between enhanced nonlinearity and manageable cooling requirements.

A related compromise appears in the superconducting detection stage. Near the CPB charge-degeneracy point, the qubit is first-order insensitive to gate-charge fluctuations while retaining a strong state-dependent energy curvature for quantum-capacitance readout \cite{Duty2005,Sillanpaa2005,Persson2010}. At the same operating point, the displacement-dependent charge coupling becomes predominantly transverse in the CPB energy basis, which allows the CNT--CPB interaction to be treated in the Jaynes--Cummings and dispersive limits \cite{Armour2002,Blencowe2004,Blais2004}. This provides a useful connection between coherent coupling and electrical detection. In the present architecture, however, quantum capacitance is not used to measure the CNT directly. Mechanical phonon information is first mapped coherently onto the CPB through the dispersive interaction, after which the CPB state is measured electrically. This separation of coherent mapping from final detection distinguishes the proposed protocol from earlier quantum-capacitance measurements of superconducting island parity \cite{Persson2010} and permits the readout resonator to remain weakly populated during the control and mapping stages that are most sensitive to backaction. The utility of this readout architecture differs from that of direct optical or cavity-based displacement detection \cite{Regal2008,Teufel2011Strong}. Its principal objective is not to maximize direct displacement sensitivity, but to convert mechanical information that depends on phonon number into a CPB phase and subsequently into a rapidly measurable electrical signal. It therefore provides access to mechanical-state information, including phonon parity and displaced-parity measurements for mechanical-state tomography, without requiring direct resolution of the CNT displacement. In terms of scalability, a movable AFM probe is less naturally parallelized than fully integrated gate or cavity architectures, but it provides spatially localized and reconfigurable actuation while separating the control, coherent mapping, and final readout functions. This combination is particularly useful for experiments that prioritize local control and quantum state characterization over larger-scale device integration.  From the perspective of condensed matter and mesoscopic physics, the CNT provides a useful setting in which several traditionally separate effects occur within the same nanoscale degree of freedom. Its mechanical spectrum can be modified by local forces and electrostatic tension, its nonlinear elasticity produces resolvable quantum anharmonicity, and its motion can couple capacitively to a mesoscopic superconducting circuit. The device therefore provides a platform for studying how nonlinear mechanical excitations, mesoscopic charge degrees of freedom, and externally controlled electromagnetic environments interact. Unlike an idealized isolated oscillator, the mechanical mode is simultaneously influenced by intrinsic damping, position fluctuations, voltage noise, cooling backaction, and superconducting readout. The ability to vary these contributions independently provides a route to controlled studies of the crossover from coherent dynamics to environmentally broadened and decohered motion \cite{Schneider2014,Remus2009,Clerk2010}.

This aspect also gives the proposed architecture broader relevance for decoherence studies and quantum information science. The AFM positioning and bias circuitry introduce experimentally adjustable noise channels whose spectra can, in principle, be measured independently and related directly to mechanical dephasing and heating. The CNT therefore offers a mesoscopic mechanical system in which the environment can be modified locally rather than treated only as an uncontrolled background. At the same time, the combination of an anharmonic mechanical spectrum with dispersive CPB coupling enables phonon number-sensitive phase accumulation, parity measurements, and displaced-parity reconstruction of the mechanical Wigner function \cite{Blais2004,Blais2021,Chang2026}. These capabilities are relevant to the preparation and verification of nonclassical mechanical states, hybrid quantum information transfer, and studies of decoherence in macroscopic collective coordinates.

The analysis also identifies several approximations that can be tested experimentally. The single-mode Duffing description assumes that higher flexural modes, internal resonances, and higher-order elastic terms remain weak. These effects can become important near strong nonlinear response or mechanical mode coupling \cite{LifshitzCross2009}. The Kerr Hamiltonian assumes weak anharmonicity relative to the carrier frequency and the validity of the rotating-wave approximation. The cooling estimates use a linearized resolved-sideband model and neglect technical heating, while the simple dispersive CNT--CPB Hamiltonian is restricted to detunings and phonon occupations for which higher-order and transition-dependent dispersive corrections remain small \cite{Boissonneault2009,Blais2021}. Finally, quantum-capacitance detection can be limited by amplifier noise, readout-induced dephasing, and nonequilibrium quasiparticles \cite{Aumentado2004,Persson2010}. These constraints are experimentally useful because they lead to distinct signatures in resonance curves, noise spectra, coherence measurements, CNT--CPB spectroscopy, and readout histograms.

The broader significance of the present framework is therefore the identification of a measurable hierarchy separating coherent control from control-induced perturbations. Rather than requiring the AFM, mechanical mode, cooling channel, and superconducting detector to be individually ideal, the analysis specifies how their characteristic rates and energy scales must compare. This provides both a design strategy and a set of experimental diagnostics for determining whether a particular device can operate in the regime of coherent, state-selective mechanical control. More generally, it illustrates how a local scanning probe can be incorporated as an active quantum control element in a hybrid mesoscopic system while its static and fluctuating backaction are treated quantitatively.

\section{Conclusions}
\label{sec:conclusion}

We have developed a unified theoretical framework for localized AFM control, nonlinear quantum dynamics, active cooling, and superconducting readout of a suspended-carbon nanotube resonator. The analysis distinguishes the coherent resonant AFM force from static force gradient, frequency noise, and force-noise backaction, connects the classical Duffing response to the Kerr anharmonicity required for state-selective mechanical control, and quantifies the cooling requirements associated with lower-frequency and electrostatically softened CNT modes. We further derived the displacement-dependent CNT--CPB interaction and showed how its dispersive limit can map mechanical phonon information, including phonon parity, onto the CPB before rapid quantum-capacitance readout. The resulting hierarchy of stability, noise, coupling, anharmonicity, cooling, and coherence requirements defines an experimentally testable operating window for the complete device.

The resulting framework provides a quantitative route for integrating localized scanning-probe actuation with a nonlinear mesoscopic mechanical system and superconducting quantum detection. Its relevance extends beyond CNT nanomechanics to condensed matter studies of coupled mesoscopic degrees of freedom, controlled investigations of decoherence and measurement backaction, and hybrid quantum information platforms in which mechanical quantum states can be prepared, manipulated, and measured with single phonon sensitivity.

\section*{Acknowledgments}
This research was funded by Tufts University Faculty Research Award.
\section*{Data Availability}
The original contributions presented in this study are included in the article. Further inquiries can be directed to the corresponding author.

\end{document}